\documentclass[aps,prx,twocolumn,amsmath,amssymb,superscriptaddress,longbibliography]{revtex4-2}
\usepackage{blindtext}
\usepackage{enumitem}
\usepackage{graphicx}
\graphicspath{{figures/}}

\usepackage{dcolumn}
\usepackage{color}
\usepackage{xcolor}
\usepackage{amsmath}
\usepackage{braket}

\usepackage{bm}
\usepackage{changes}
\usepackage{chemformula}

\usepackage[breaklinks,colorlinks,bookmarks=false,citecolor=blue,linkcolor=red,urlcolor=blue]{hyperref}
\usepackage{cleveref}
\crefname{equation}{Eq.}{Eqs.}
\Crefname{equation}{Eq.}{Eqs.}
\crefname{figure}{Fig.}{Figs.}
\Crefname{figure}{Fig.}{Figs.}
\crefname{section}{Sec.}{Secs.}
\Crefname{section}{Sec.}{Secs.}

\definecolor{darkgreen}{rgb}{0.0, 0.44, 0.1}

\usepackage{changes}

\begin{document}

\title{Mott insulating states with competing orders in the triangular lattice Hubbard model}

\author{Alexander Wietek}
\email{awietek@flatironinstitute.org}
\affiliation{Center for Computational Quantum Physics, Flatiron Institute, 162 Fifth Avenue, New York, NY 10010, USA}
\author{Riccardo Rossi}
\affiliation{Center for Computational Quantum Physics, Flatiron Institute, 162 Fifth Avenue, New York, NY 10010, USA}
\affiliation{Institute of Physics, \'Ecole Polytechnique F\'ed\'erale de Lausanne (EPFL), CH-1015 Lausanne, Switzerland}
\author{Fedor \v{S}imkovic IV}
\affiliation{CPHT, CNRS, {\'E}cole Polytechnique, IP Paris, F-91128 Palaiseau, France}
\affiliation{Coll{\`e}ge de France, 11 place Marcelin Berthelot, 75005 Paris, France}
\author{Marcel Klett}
\affiliation{Max-Planck-Institut f\"ur Festk\"orperforschung, Heisenbergstraße 1, 70569 Stuttgart, Germany}
\author{Philipp Hansmann}
\affiliation{Department of Physics, University of Erlangen-N{\"u}rnberg, 91058, Erlangen, Germany}
\author{Michel Ferrero}
\affiliation{CPHT, CNRS, {\'E}cole Polytechnique, IP Paris, F-91128 Palaiseau, France}
\affiliation{Coll{\`e}ge de France, 11 place Marcelin Berthelot, 75005 Paris, France}
\author{E. Miles Stoudenmire}
\affiliation{Center for Computational Quantum Physics, Flatiron Institute, 162 Fifth Avenue, New York, NY 10010, USA}
\author{Thomas Sch\"afer}
\affiliation{Max-Planck-Institut f\"ur Festk\"orperforschung, Heisenbergstraße 1, 70569 Stuttgart, Germany}
\author{Antoine Georges}
\affiliation{Coll{\`e}ge de France, 11 place Marcelin Berthelot, 75005 Paris, France}
\affiliation{Center for Computational Quantum Physics, Flatiron Institute, 162 Fifth Avenue, New York, NY 10010, USA}
\affiliation{CPHT, CNRS, {\'E}cole Polytechnique, IP Paris, F-91128 Palaiseau, France}
\affiliation{DQMP, Universit{\'e} de Gen{\`e}ve, 24 quai Ernest Ansermet, CH-1211 Gen{\`e}ve, Suisse}

\date{\today}

\begin{abstract}
The physics of the triangular lattice Hubbard model exhibits a rich
phenomenology, ranging from a metal-insulator transition,
intriguing thermodynamic behavior, and a putative spin liquid phase at
intermediate coupling, ultimately becoming a magnetic insulator at strong coupling. 
In this multi-method study, we combine a finite-temperature
tensor network method, minimally entangled thermal typical states (METTS), 
with two Green function-based methods, connected-determinant diagrammatic Monte Carlo (DiagMC) and
cellular dynamical mean-field theory (CDMFT), to establish several aspects of
this model. We elucidate the evolution from the metallic to the insulating
regime from the complementary perspectives brought by these different methods. 
We compute the full thermodynamics of the model on a width-4 cylinder using 
METTS in the intermediate to strong  coupling regime. 
We find that the insulating state hosts a large entropy at intermediate temperatures, 
which increases with the strength of the coupling. 
Correspondingly, and consistently with a thermodynamic Maxwell relation, 
the double occupancy has a minimum as a function of temperature which 
is the manifestation of the Pomeranchuk effect of increased localisation upon heating. 
The intermediate coupling regime is found to exhibit both pronounced chiral 
as well as stripy antiferromagnetic spin correlations. We propose a scenario  
in which time-reversal symmetry broken states compete with stripy-spin states at lowest temperatures. 
\end{abstract}

\maketitle


\section{Introduction}
\label{sec:introduction}

The interplay between strong electronic interactions and geometric frustration gives rise to a plethora of 
intriguing physical phenomena. It also raises fundamental questions that are still largely open, such as how insulating 
spin liquids transition into a metallic or superconducting phase upon reducing the interaction strength or introducing doped charge carriers. 

Several classes of experimental platforms are available in which these questions can be explored. 
The most recent one is the rapidly developing field of twisted moir\'{e}  heterostructures  of two-dimensional materials, such as 
graphene~\cite{Cao2018a,Cao2018b,Yankowitz2019} or transition-metal dichalcogenides~\cite{Wu2018,Wu2019}. 
Recent work has demonstrated that these heterostructures provide a versatile platform for quantum materials design in which a broad range 
of lattice and band structures can be engineered~\cite{Kennes_simulator,Wang_Dean_2020}. 
A triangular lattice structure, which is the focus of the present paper, can be realized in this context for B 
moir\'e superlattices~\cite{Tang_2020,Regan_2020}, 
twisted WSe$_2$ double bilayers~\cite{An_2020} as well as twisted bilayer 
Boron Nitride~\cite{Xian_2019,Ni_2019}. 
The observation of a Mott insulating state in e.g. the WSe$_2$/WS$_2$ moir\'e superlattice system~\cite{Tang_2020} provides 
direct experimental evidence of the importance of strong electronic correlations in these materials. 
We also note that the triangular superlattice dichalcogenide 1T-TaS$_2$ has been proposed to host a spin-liquid state~\cite{Law2017,Kratochvilova2017,Qiao2017,Murayama2020}.

Besides moir\'{e} materials, strong electronic correlations in the context of (anisotropic) triangular lattice structures 
are also directly relevant to the two-dimensional molecular materials of the $\kappa$-ET 
family~\footnote{ET, sometimes also abbreviated as BEDT-TTF, stands for bis(ethylene-dithio)
tetrathiafulvalene.}. This class of materials has been the subject of 
intense experimental research and displays a diversity of remarkable phenomena 
(for reviews, see e.g.\cite{Powell2011,Kanoda2011}). 
Among those are Mott insulating phases with 
either magnetic long-range order or spin liquid behavior as in e.g. \ch{$\kappa$-(ET)2Cu2(CN)3}, 
a pressure-induced metal-insulator transition (MIT). Several experiments found evidence
of first-order phase transitions at finite-temperature up to a proposed critical temperature of $\sim 20$~K~\cite{Furukawa2015,Kagawa2005}. Moreover, superconductivity with a critical temperature reaching $\sim 14$~K has been found. 
~\cite{Komatsu1996,Lefebvre2000,Limelette2003,Shimizu2003,Kagawa2005,Kurosaki2005,Ohira2006,Itou2007,Kandpal2009,Maegawa2011,Isono2014}. 
Finally, transition metal oxides such as the layered superconductor Li$_x$NbO$_2$ also form triangular lattices, 
with structural similarities 
to some of the dichalcogenides~\cite{Pickett2006,Lee2007,Soma2020}.

While the Hubbard model \cite{Hubbard1963, Hubbard1964, Gutzwiller1963, Kanamori1963} on the triangular lattice  is directly relevant to this wide variety
of materials, it is also a paradigmatic model of strongly correlated electrons 
subject to geometric frustration and has therefore been subject to intense
computational and theoretical research \cite{Qin2021,Arovas2021}. 
However, due to the high complexity of the problem,  
only a partial understanding of its physics has been reached. 
The model is defined by the Hamiltonian: 
\begin{equation}
  \label{eq:hubbardmodel}
  \hat{H} = -t \sum_{\langle i, j \rangle, \sigma}
  \left( \hat{c}^\dagger_{i\sigma} \hat{c}_{j\sigma}  + \hat{c}^\dagger_{j\sigma} \hat{c}_{i\sigma} \right)
  + U \sum_{i}\hat{n}_{i\uparrow} \hat{n}_{i\downarrow},
\end{equation}
where $\hat{c}^\dagger_{i\sigma}, \hat{c}_{i\sigma}$ denote the fermionic
creation and annihilation operators on site $i$ with fermion spin $\sigma$, 
$\hat{n}_{i\sigma} = \hat{c}^\dagger_{i\sigma} \hat{c}_{i\sigma}$, and $\braket{i,j}$ denotes
summation over nearest-neighbor bonds of the triangular lattice. 

At half-filling ($\langle \hat{n}_{i\uparrow}+\hat{n}_{i\downarrow}\rangle = 1$) 
the model has a metallic phase for small $U/t$, while it is an insulator with 
long-range magnetic order in the large $U/t$ limit for $T=0$. At finite temperatures true long-range magnetic order is prohibited in two dimensions by the Mermin-Wagner theorem \cite{Mermin1966, Hohenberg1967}.
It has been suggested early on that an intermediate insulating phase without magnetic long-range order exists between 
these two phases, at intermediate $U/t$~\cite{Morita2002}\footnote{A narrow region of superconductivity has also been suggested using CDMFT between the metal and the non-magnetic insulator~\cite{Kyung2006}.}.
%
The existence of this intermediate phase has been corroborated by several different computational
methods~\cite{Kyung2006,Sahebsara2008,Laubach2015,Misumi2017,Tocchio2008,Yoshioka2009,Yang2010,Szasz2020}.
Recent density matrix renormalization group (DMRG) studies showed strong
evidence that the intermediate phase ground state realizes a gapped chiral spin
liquid (CSL)~\cite{Szasz2020,Szasz2021,Chen2021}.
This elusive state of matter has been proposed in the late
1980s~\cite{Kalmeyer1987,Wen1989} and in the last years has been found to be
stabilized in several frustrated spin systems~\cite{Messio2012,Schroeter2007,He2014,Gong2014,Wietek2015,Bieri2015,Bauer2014}, 
including extended triangular lattice spin-$1/2$ Heisenberg models~\cite{Wietek2017,Gong2017}. 
The demonstration of the emergence
of topological superconductivity upon hole-doping the triangular lattice
CSL~\cite{Jiang2020} (see also \cite{Zhu2020}) constitutes an exciting prospect
for the physics of the moir\'{e} materials and organic superconductors.
Other previous suggestions on the nature of the intermediate phase include a
Gutzwiller projected Fermi sea~\cite{Motrunich2005,Sheng2009,Block2011} and
other forms of gapless spin liquids~\cite{Grover2010,Tocchio2013,Mishmash2013}.
However, the existence of the intermediate CSL phase is also challenged by earlier DMRG results~\cite{Shirakawa2017} as well as recent variational Monte Carlo studies, suggesting the absence of a spin liquid phase close to the metal-insulator transition~\cite{Tocchio2020,Tocchio2021}. Other recent DMRG results on width 3 cylinders have suggested a gapless spin liquid being realized in the intermediate phase~\cite{Peng2021}.

Computational methods for studying quantum many-body systems rely on diverse concepts 
and usually involve approximations whose validity has to be subject to critical evaluation. 
In the context of the Hubbard model, direct comparisons
and benchmark studies involving multiple methods have proved successful in establishing the physics 
in the strongly correlated~\cite{LeBlanc2015,Zheng2017,Qin2020} and intermediate coupling~\cite{Schaefer2020} 
regimes beyond the uncertainties associated with the limitations of one particular method (see also a recent study on the kagome lattice \cite{Kaufmann2021}). 
This multi-method approach is currently playing a crucial role in the field and accelerates further theoretical and computational developments. 

In this article, we combine conceptually different methods to investigate the physics of the 
half-filled triangular lattice Hubbard model at finite temperature. 
For some physical observables, the results from these different methods can be directly compared, 
but each method also comes with physical observables that it can more naturally address. 
Such a `multi-method, multi-messenger' approach~\cite{Schaefer2020} therefore allows us 
to investigate the physics of this complex model from different perspectives. 

On the one hand, we employ the minimally entangled thermal typical state method
(METTS)~\cite{White2009,Stoudenmire2010} which is an extension of DMRG to finite temperature. 
On the other hand, we use two Green function based techniques, 
the diagrammatic Monte Carlo method (DiagMC~\cite{Prokofev1998}) 
in its connected determinant version~\cite{Rossi2017,Simkovic2019, Moutenet2018, Rossi2020, MCMCMC}, 
dynamical mean-field theory (DMFT \cite{Metzner1989,Georges1992,Georges1996,Rubtsov2004}) 
and a cluster extension \cite{Maier2005} thereof: cellular DMFT (CDMFT)~\cite{Kotliar2001,Lichtenstein2000} 
in its center-focused formulation~\cite{Klett2020}. 
Such a `handshake' between wave-function based and Green function based methods is a 
notable advance which opens new perspectives for the study of quantum many-body problems at finite temperature. 

Each of these methods has strengths and limitations which we now briefly describe. 
As a matrix-product state technique METTS can be applied with
high precision on cylindrical geometries of finite circumference,  
as demonstrated recently by some of the authors in the case of the hole-doped square
lattice Hubbard model~\cite{Wietek2020}, where a detailed description of the
implementation of the method can be found. In this manuscript we mostly focus on a 
particular cylindrical geometry of circumference 4, called the YC4 geometry 
shown in \cref{fig:snapshots}. Selected results on the XC4 and YC3 geometries (see e.g. \cite{Szasz2020}) are also presented for comparison. Since we demonstrate convergence of
our results in the maximal bond dimension $D_{\max}$ (\Cref{app:metts}), 
the main limitation is in the finite transverse size. 

As the triangular-lattice Hubbard model is afflicted by the fermionic sign problem, we cannot use traditional Quantum Monte Carlo techniques~\cite{TroyerSignProblem}. Diagrammatic Monte Carlo can work directly in the thermodynamic limit and is therefore immune from the sign problem, while being numerically exact: it is possible to compute quantities with arbitrary precision given enough computational time, and the convergence can be checked by comparing results from different expansion orders. Reaching the strong-coupling regime can, however, be 
hindered by the increased difficulty of resumming the perturbative series beyond 
their radius of convergence \cite{Simkovic2019}, and many orders must be computed, which in itself can present computational challenges. In this work, using the recent computational advances of the connected-determinant version~\cite{Rossi2017}, we are able to compute up to 10 orders of the perturbative expansion at fixed density; this is achieved~\cite{RossiSimkovic} by renormalizing the chemical potential in the spirit of~\cite{Rossi2020}. Thanks to the high orders reached, we manage to get converged results with controlled errorbars at temperatures $T/t=0.1$ and up to $U/t=10$.

CDMFT also works directly in the thermodynamic limit for the lattice, but it retains only a finite number of 
real-space components of the self-energy organized according to spatial locality and approximated by their 
value on a (self-consistent) cluster of finite size $N_c$. 
The method is controlled in the sense that it converges to the exact solution in the limit $N_c\!\rightarrow\!\infty$, but 
in practice, this convergence can only be reached in specific parameter regimes. 
Here, CDMFT is used in a twofold way: (i) as an approximation with $N_c\!=\!7$ restricted to the paramagnetic (PM) phase [CDMFT-7 (PM)] and (ii) as an approximation with $N_c\!=\!4$ (CDMFT-4) allowing for magnetic ordering. Unless noted otherwise the label ``CDMFT" denotes results from the first variant. In either case on-site and nearest-neighbor components of the self-energy are taken into account, besides all temporal (quantum) correlations already present in single-site DMFT.


This article is organized as follows. We discuss the transition from 
a metallic state at weak-coupling to an insulating state at strong coupling
in \cref{sec:mitransition}. There, we perform a critical comparison between
our numerical methods and propose that, in the accessible range of temperatures, 
a smooth crossover between these states is found rather than 
a first-order phase transition. 
In \cref{sec:selfenergy} we discuss the locality of the electronic self-energy 
by comparing results from CDMFT and DiagMC.
In \cref{sec:thermodynamics} we investigate the basic thermodynamic
properties of the system and firmly establish an order-by-disorder effect,
where increasing temperature decreases the double occupancy. We relate this
effect to an increase in entropy upon increasing the interaction strength
via a Maxwell relation. \Cref{sec:magnetism} discusses competing (magnetic)
orders as a function of temperature and interaction strength. In particular,
we investigate magnetic structure factors and the chiral susceptibility to 
propose a scenario where chiral and stripy antiferromagnetic spin correlations coexist at low temperatures. Finally, we summarize and discuss our findings in \cref{sec:discussion}.

\section{Metal-Insulator crossover}
\label{sec:mitransition}

\begin{figure}[t]
    \centering
    \includegraphics[width=\columnwidth]{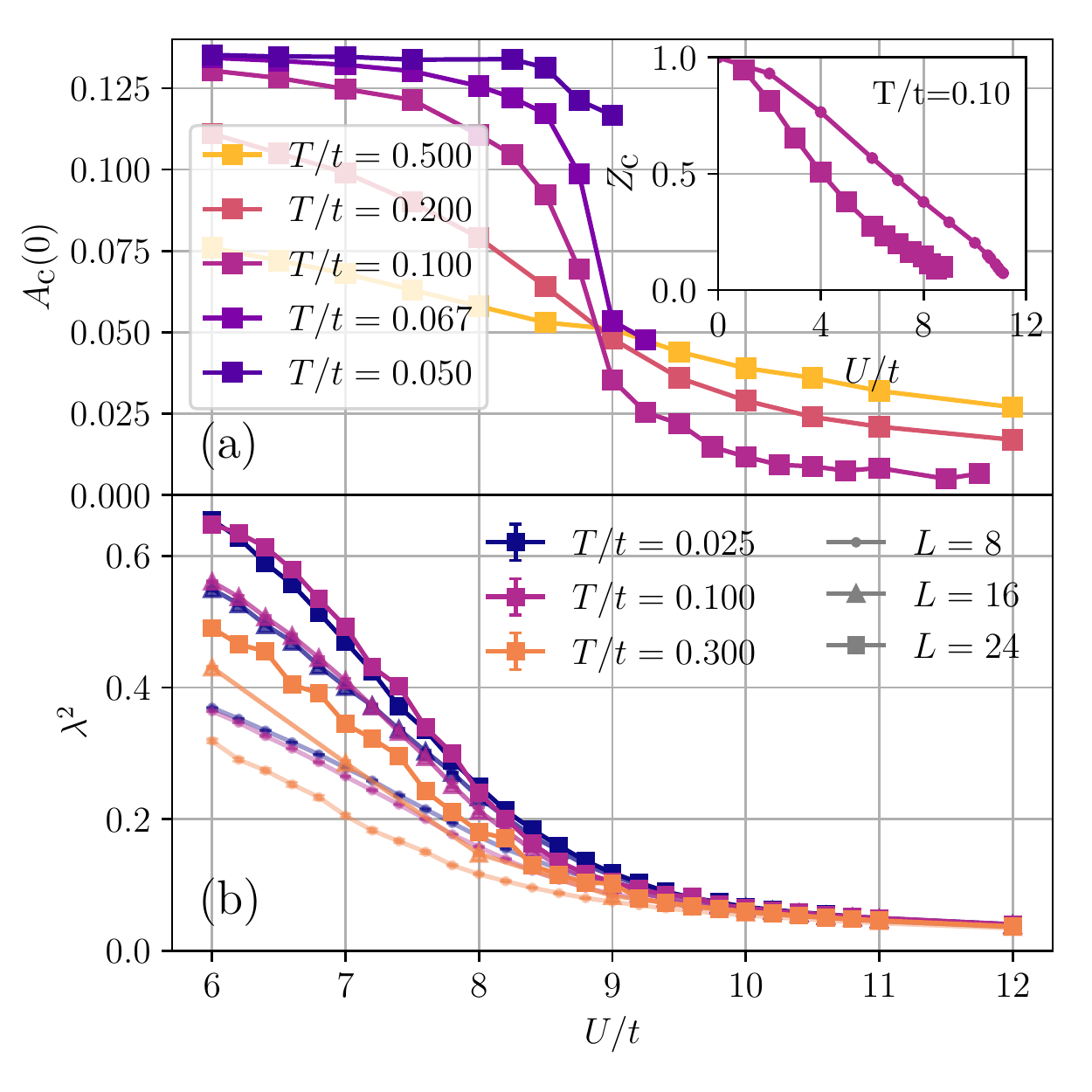}
    \caption{Metal-Insulator crossover at finite-temperature from CDMFT 
    and METTS. (a) Spectral weight at the Fermi level from CDMFT. The inset shows the quasiparticle renormalization factor for the central site in
    CDMFT (squares) as well as for PM-restricted DMFT (dots). A
    drop of the spectral weight at the Fermi level is observed.
    (b) Normalized localization length as a function of temperature
    and system size as obtained from METTS on the YC4 cylinder. 
    Simulations have been performed with maximal bond dimension $D_{\max}=4000$.
    The normalized localization length attains a finite value in the
    insulating regime and increases in the linear system size $L$ in 
    the metallic regime, where higher temperatures decrease the 
    localization length.}
    \label{fig:mit_transition}
\end{figure}

We begin by investigating the evolution from a metallic state at weak coupling to a Mott 
insulating state at strong coupling. At high enough temperature, this is a crossover. 
Whether it remains a crossover down to lowest temperatures or whether a phase transition also exists at low but finite temperature is discussed at the end of this section. 

In order to identify this crossover, we consider two complementary observables, which are accessible within the CDMFT and 
METTS frameworks respectively. 
The first one is the zero-frequency value of the local (on-site) electronic spectral function:
\begin{equation}
    A_{\text{c}}(\omega=0)=-\frac{1}{\pi}\,\text{Im}G_{\text{c}}(i\omega_n\rightarrow i0^+).
\end{equation}
This quantity is evaluated by considering the central site of the cluster within the center-focused CDMFT method (see App.~\ref{app:cdmft} and \cite{Klett2020}) --- 
hence the subscript in $A_c$. The extrapolation to zero frequency is obtained from a fit of the Matsubara frequency Green's function 
$G(i\omega_n)$. We have also calculated within CDMFT the local and nearest-neighbor components of the self-energy 
and can extract the low-frequency slope:
\begin{equation}
    Z_c\,=\,\left[1-\frac{\partial\Sigma_c}{\partial\omega}\bigg|_{\omega\rightarrow 0}
\right]^{-1},
\end{equation}
which is also a good indicator of the MIT. The non-local components of the self-energy are found to be quite small for weak to intermediate $U/t$ (see Sec.~\ref{sec:selfenergy} for more details), hence $Z_c$ is a reasonable approximation in this regime 
to the spectral weight of quasiparticles. 
$A_c(0)$ is plotted in Fig.~\ref{fig:mit_transition}(a) as a function of 
$U/t$ for several values of temperature, along with 
$Z_c$ at $T/t=0.1$ (inset). We see that for each value of $T$, $A_c(0)$ 
undergoes a marked drop as $U/t$ is increased, signalling 
a crossover from a metal with a large value of the zero-frequency spectral
density to an insulator with a small one 
(but as expected still finite at non-zero $T$). 
Being a crossover there is a certain arbitrariness in defining its location
precisely but it is apparent that, at the lowest 
temperatures, it occurs for $8\lesssim U/t \lesssim 9$. Correspondingly, $Z_c$
drops rapidly as $U/t$ is increased. 
At still larger values of $U/t$, the CDMFT self-energy has the characteristic
divergent low-frequency behavior of an insulator, see 
Fig.~\ref{fig:sigma_CDMFT_iwn} in Appendix~\ref{app:cdmft}. 
Importantly, we see that $A_c(0)$ increases upon cooling for $U/t\lesssim 9$, 
while it decreases upon cooling for $U/t\gtrsim 9$, which are the expected behaviors 
in a metallic and an insulating regime, respectively. We note that for small temperatures and large interactions the CDMFT calculation becomes increasingly difficult due to the fermionic sign problem.

The evolution from a metallic state to an insulating state can also be
characterized by the localization of electrons~\cite{Kohn1964,Resta1999}.
An appropriate measure of localization is given by the localization length
$\lambda$, which on open boundary conditions as employed by METTS is 
defined as~\cite{Resta2018,Resta2019,Resta2020},
\begin{equation}
    \label{eq:localizationlength}
    \lambda^2 = \frac{1}{N}\left(\braket{X^2} - \braket{X}^2\right).
\end{equation}
Here, $X = \sum_i r_i n_i$ denotes the position operator, where $r_i$
denote the coordinates of the lattice, $n_i$ the local density operators,
and $N$ the total number of sites.

At zero temperature, the localization length $\lambda$ is directly related
to the real part of the conductivity $\sigma(\omega)$ by~\cite{Souza2000}
\begin{equation}
    \label{eq:swmintegral}
    \lambda^2 = \frac{\hbar}{\pi e^2 n} 
    \int_0^\infty \frac{d\omega}{\omega}\; \text{Re} \; \sigma(\omega), 
    \quad (T=0)
\end{equation}
where $e$ denotes the electron charge and $n$ the average density. 
The integral on the right-hand side, also referred to as Souza-Wilkens-Martin
integral~\cite{Souza2000,Resta2018,Resta2020}, diverges in the metallic
regime and attains a finite value in the insulating regime for
$N\rightarrow\infty$. The behavior
of $\lambda^2$ for temperatures $T/t= 0.025, 0.100, 0.300$ and YC4 
cylinder lengths $L=8, 16, 24$ computed by METTS is shown in
\cref{fig:mit_transition}(b). The metallic and insulating regimes can be coarsely
distinguished by the behavior of $\lambda^2$. Whereas in the insulating
regime, $\lambda^2$ is almost constant as a function of the cylinder length
$L$ and temperature, it increases with $L$ in the metallic regime. We also 
observe that at higher temperature, such as $T/t=0.300$ in
\cref{fig:mit_transition}(b), the localization length decreases, indicating
increased localization of the system. 

\begin{figure}[t]
    \centering
    \includegraphics[width=\columnwidth]{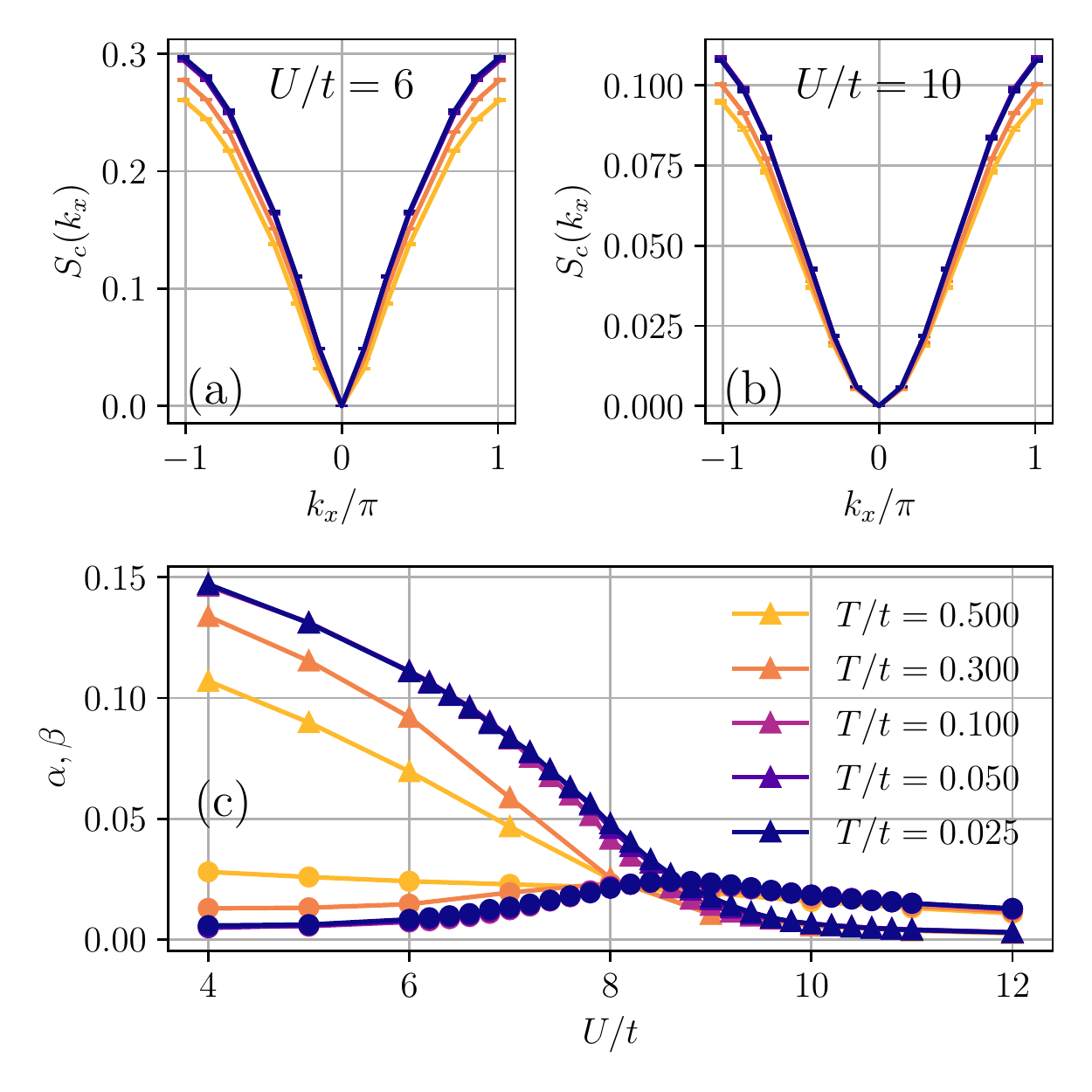}
    \caption{Static charge structure factor $S_c(\mathbf{k})$ for $k_y=0$ at  various temperatures (see legend in (c)) on the $16\times 4$ YC4 cylinder from METTS with
    maximal bond dimension $D_{\max}=4000$. 
    (a) In the metallic regime at $U/t=6$, the density structure factor behaves
    as $S_c(k_x) \approx \alpha|k_x|$. (b) In the insulating regime at $U/t=10$
    we observe $S_c(k_x) \approx \beta k_x^2$. (c) Optimal fit parameters
    $\alpha,\beta$ for the ansatz $S_c(k_x) = \alpha |k_x| + \beta k_x^2$ of 
    $S_c(k_x)$ close to $k_x=0$. $\alpha$ (resp. $\beta$) is shown as triangles
    (resp. circles). The crossover interaction strength $U_c/t$ can be defined by
    the intersection of $\alpha$ and $\beta$. We observe $U_c/t$ shifting towards
    weaker interactions at higher temperatures.}
    \label{fig:dofk}
\end{figure}

Furthermore, we study the behavior of the charge structure factor given by,
\begin{equation}
  \label{eq:chargestructure}
  S_{\textrm{c}}(\bm{k}) = \frac{1}{N}\sum_{l,m=1}^N
  \text{e}^{i\bm{k}\cdot(\bm{r}_l - \bm{r}_m)}
  \braket{(n_l - \braket{n_l}) (n_m - \braket{n_m})},
\end{equation}
where $n_l$ denotes the local density at site $l$. When using METTS we
are working in the canonical ensemble with zero global charge fluctuation,
which implies $S_{\textrm{c}}(0)=0$ at any temperature. The behavior 
of $S_{\textrm{c}}(\bm{k})$ around $\bm{k}=0$ is indicative of metallic
or insulating behavior. While a metallic state at $T=0$ is characterized
by a linear charge dispersion~\cite{Feynman1954,Capello2005,Tocchio2011,DeFranco2018},
\begin{equation}
  S_{\textrm{c}}(\bm{k}) \approx \alpha |k_x|,
\end{equation}
an insulating state exhibits a quadratic dispersion,
\begin{equation}
  S_{\textrm{c}}(\bm{k}) \approx \beta k_x^2.
\end{equation}
The behavior of the charge structure factor on the $16\times 4$ YC4 cylinder
from METTS at various temperatures in \cref{fig:dofk}. We compare between
the metallic regime at $U/t=6$ in (a) and the insulating regime at $U/t=10$ in 
(b). The two regimes clearly exhibit the expected linear (resp. quadratic)
behavior close to $\bm{k}$. We observe a rather mild temperature dependence.
To quantify this behavior, we make the following ansatz close to $k_x=0$,
\begin{equation}
\label{eq:dofkansatz}   
  \tilde{S}_{\textrm{c}}(\bm{k}) = \alpha |k_x| + \beta k_x^2.
\end{equation}
The value of $\alpha$ (resp. $\beta$) can be interpreted as a metallic
(resp. insulating) weight. We fit this ansatz to the numerical data shown in
\cref{fig:dofk}(a,b) for the seven $k$-points closest to the origin. Results
for a range from $U/t=4$ to $U/t=12$ are shown in \cref{fig:dofk}(c). 
At low $U/t$ in the metallic regime, we observe that increasing temperature
decreases the metallic weight $\alpha$ while increasing the insulating weight. 
In the insulating regime, we observe only a weak temperature dependence of 
$\alpha$ and $\beta$. We can define a crossover interaction strength $U_c/t$
by the intersection of $\alpha(U)$ and $\beta(U)$. We observe that $U_c/t$ shifts
towards weaker interaction strengths when increasing temperatures. From this we estimate $U_c/t\approx 8.7$ at $T/t=0.025$,  $U_c/t\approx 8.5$ at $T/t=0.100$,
and $U_c/t\approx 8.0$ at $T/t=0.500$. 
This also implies that, for a fixed $U/t$ in that range, the system undergoes increased localisation 
upon heating.

\begin{figure}[t!]
    \centering
 \includegraphics[width=\columnwidth]{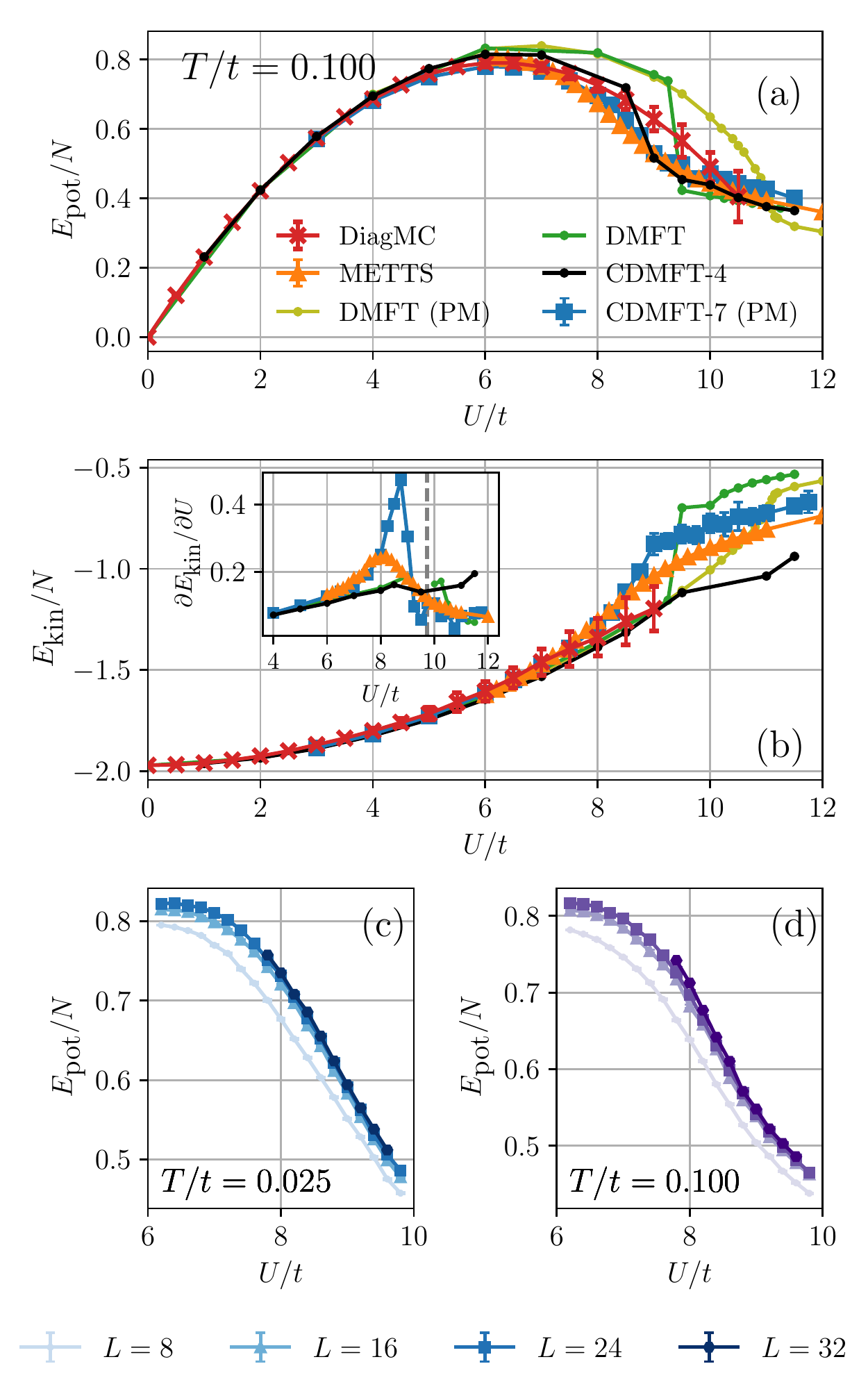}
    \caption{Comparison of energies from different computational methods at
    $T/t=0.1$. METTS results are obtained on a $16\times 4$ YC4 cylinder with
    $D_{\max}=4000$,
    cluster DMFT is performed on a $7$-site cluster [restricted to its paramagnetic solution, CDMFT-7 (PM)] and on a $4$-site cluster (allowing for spin-symmetry breaking, CDMFT-4), while DiagMC is 
    numerically-exact within the estimated errorbar. 
    (a) Potential energy $E_{\textrm{pot}}$. METTS and cluster-DMFT agree
    with the DiagMC results up to $U/t\approx 7.5$. Beyond this point we find
    excellent agreements between cluster DMFT and METTS. 
    (b) Kinetic energy $E_{\textrm{pot}}$. We observe good agreement between all
    methods up to $U/t = 8$. The inset shows the derivative of the kinetic energy w.r.t. the coupling strength $U$. (c,d) The potential energy
    density from METTS as a function of cylinder length $L$ at temperatures
    $T/t=0.025$ and $T/t=0.100$. We observe a smooth dependence on the coupling 
    strength and for both temperatures upon increasing the system size.}
    \label{fig:comparison}
\end{figure}
To further study the metal to insulator crossover we investigate the potential energy,
\begin{equation}
    \label{eq:potentialenergy}  
    E_{\textrm{pot}} = U\sum_{i}\braket{\hat{n}_{i\uparrow} \hat{n}_{i\downarrow}},
\end{equation}
and the kinetic energy,
\begin{equation}
    \label{eq:kineticenergy}  
    E_{\textrm{kin}} = -t  \sum_{\langle i, j \rangle, \sigma}
  \langle \hat{c}^\dagger_{i\sigma} \hat{c}_{j\sigma} +  \hat{c}^\dagger_{j\sigma} \hat{c}_{i\sigma} \rangle.
\end{equation}
Since these quantities are accessible with all our methods, we show a direct
comparison in \cref{fig:comparison}(a,b) to assess the effects of finite cluster size 
in CDMFT and finite cylinder size in METTS. We focus on a temperature of $T/t=0.1$,
for which we perform numerically-exact simulations with DiagMC up to $U/t=10.5$, which
are used as benchmark. Remarkably, results from all methods agree within
error bars up to an interaction strength of $U/t=8$. Beyond this point, we still
observe that the potential energy from METTS compares well with CDMFT [with $N_c\!=\!7$ and restricted to its paramagnetic solution, CDMFT-7 (PM)] up to the
strong coupling regime $U/t=12$. A key difference between METTS and CDMFT is 
seen in the kinetic energy, which is lower for METTS in the strong coupling regime.
We also note that the CDMFT kinetic energy exhibits a slope discontinuity around $U/t \sim 9$
which is not observed by METTS: as discussed below, this is due to the fact that 
a finite-temperature metal-insulator 
transition is found within CDMFT (as well as DMFT) when constrained to non-magnetic solutions. 
Nonetheless, the data from CDMFT overall agrees more closely with METTS than the data from single-site DMFT.

Above $T/t\simeq 0.1$, all methods agree that the passage from the metal to the insulator is a smooth crossover~\cite{Shaozhi2020}. 
Furthermore, it is clear from 
previous work~\cite{Kyung2006,Sahebsara2008,Laubach2015,Goto2016,Misumi2017,Szasz2020,Szasz2021},
that at $T=0$ a metal-insulator phase transition (MIT) takes place. 
We now discuss whether our data allow us to settle whether a sharp MIT also exists at low 
but finite temperature or whether a smooth crossover applies for any non-zero temperature. 

Let us first recall what the situation is 
in the single-site DMFT approximation. When solving the DMFT equations constrained 
to solutions without long-range magnetic order, one indeed finds that a first-order
MIT develops for $T<T_c^{\text{DMFT}}\simeq 0.1 t$ ($U_c^{\text{DMFT}}/t\simeq 11$) as 
previously demonstrated by several
authors and also shown in Fig.~\ref{fig:dmft_magnetic} of Appendix~\ref{app:dmft}. 
Since there is no symmetry distinction at finite temperature between a metal and an
insulator with no broken symmetries, a first-order transition line ending at a
second-order critical endpoint $(U_c,T_c)$ (analogous 
to a liquid-gas transition) is a priori possible. This is what happens 
in DMFT~\cite{Georges1993,Rozenberg1994,Georges1996,Kotliar1999,Rozenberg1999,
Kotliar2000,Bulla2001,Limelette2003_Science,Liebsch2009,vanLoon2020}, 
as well as in cluster extensions of 
DMFT~\cite{Parcollet2004,Dang2015,Lee2008,Ohashi2008,Semon2012,Vranic2020} 
when restricted to non-magnetic solutions. This 
transition is the reason for the cusp in the kinetic energy found with these methods, 
as apparent on Fig.~\ref{fig:comparison}(b) around $U_c^{\text{CDMFT}}/t\simeq 9$. 
However it should be emphasized that, when allowing for spin and translational
symmetry breaking, the single-site DMFT approximation yields a solution with
$120^\circ$ N\'eel ordering for $U/t\gtrsim 9.5$~\cite{Goto2016} (see Appendix~\ref{app:dmft}): 
this is the true minimum of the 
free-energy in the DMFT approximation, hence overshadowing the first-order
non-magnetic MIT. Incidentally, we note that the magnetically ordered DMFT solution
yields a rather good approximation to both the kinetic and potential energy
(Fig.~\ref{fig:comparison}).

Analogously to the DMFT case, we also computed the energetics within CDMFT on a $N_c\!=\!4$ site cluster, however now allowing for magnetic symmetry breaking (CDMFT-4). The order sets in at similar interaction strengths as in the case of DMFT. Due to the smaller size of the cluster, and therefore the shorter non-local correlations that are included in the calculation, deviations from CDMFT-7 (PM) appear for the potential energy in the paramagnetic regime at intermediate coupling. At larger interaction strengths, when CDMFT-4 has ordered with a $120^\circ$ N\'eel pattern, the potential energy acquires similar values to both METTS and CDMFT-7 (PM). In case of the kinetic energy, we observe deviations also in the ordered phase. They might root in the fact that the cluster geometries differ between CDMFT-7 (PM) and CDMFT-4 (for further details we refer to App.~\ref{app:cdmft}). We note that calculations in the symmetry-broken phase for a $N_c\!=\!7$ site cluster are not feasible at the moment due to the fermionic sign problem.

Of course, the magnetic solution at non-zero temperature is an artefact of the
mean-field approximation inherent to DMFT, and one may argue that because 
fluctuations and Mermin-Wagner theorem actually prevent ordering, the existence 
of a finite-$T$ MIT in paramagnetic DMFT/CDMFT is a hint that a similar phenomenon 
might take place in our model. On a qualitative level, frustration appears as a
favorable factor by further suppressing ordering. 
The ET-organic materials  with an (anisotropic) triangular structure do
display such a transition experimentally~\cite{Kanoda2011}. 

Our METTS and DiagMC results do not provide evidence for such a first-order MIT or liquid-gas 
critical endpoint at finite temperature. In the range of temperatures that we could investigate,  
the kinetic and potential energy displayed in Fig.~\ref{fig:comparison} do not appear to have a singularity as a function of $U/t$. 
However, we acknowledge that limitations of our computational methods prevent us to reach a 
definitive conclusion about this issue.
Our METTS results yield a smooth crossover between the metallic and the
insulating regime for temperatures down to $T/t=0.025$. This is expected,
since our simulations are performed on a finite system. Hence, observables
will depend smoothly on the model parameters. We have, however, investigated
the possibility of a discontinuity, indicative of a first-order phase transition,
developing as a function of the length of the YC4 cylinder. Our results are
shown for temperatures $T/t=0.025$ in \cref{fig:comparison}(c) and for
$T/t=0.100$ in \cref{fig:comparison}(d). The smooth behavior at all system sizes does not
exhibit any tendency to develop a discontinuity for $L\rightarrow \infty$.
However, it is possible that the chosen cluster geometry or the finite precision we achieve conceal potential singularities 
developing in the infinite-volume limit. The data points shown in
\cref{fig:mit_transition} are spaced by $\Delta U= 0.2t$ where the maximal
absolute statistical error is of size
$\varepsilon \approx 5 \cdot 10^{-3}$. 
The DiagMC results, while dealing with the infinite system, are limited in the present work to 
$T/t\gtrsim 0.1$ and $U/t\lesssim 10$. This is largely due to difficulties in computing enough expansion coefficients with small enough error bars to allow for controlled resummations of the perturbative series beyond the aforementioned values of $U$. 

\begin{figure}[t]
    \centering
    \includegraphics[width=\columnwidth]{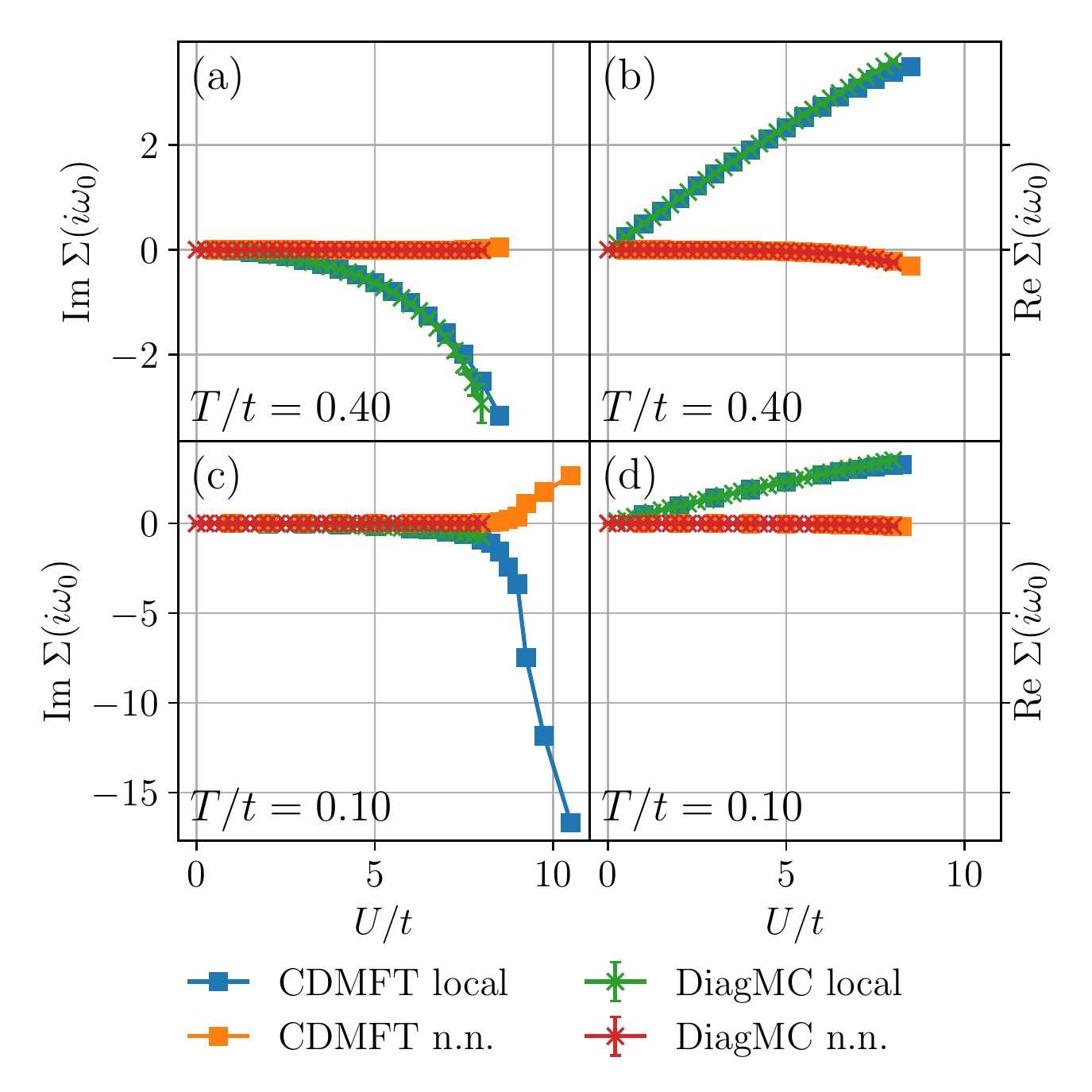}
    \caption{Imaginary (left panels) and real (right panels) parts of the local (blue) and nearest neighbor (orange) self-energy at its lowest Matsubara frequency calculated by DiagMC (crosses) and CDMFT (squares) as a function of $U/t$ for two different temperatures $T/t=0.4$ (upper panels) and $T/t=0.1$ (lower panels). A substantial increase of  non-local correlations is observed at low temperatures, however, at interactions ($U/t \approx 9.25$) larger than the one of increased local correlations ($U/t \approx 8$).}
    \label{fig:self_energy}
\end{figure}

\section{(Non-)locality of correlations: self-energies}
\label{sec:selfenergy}

Although the electronic Coulomb interaction is modelled as a purely local repulsion in the Hubbard model in Eq.~(\ref{eq:hubbardmodel}), the correlations it generates can be non-local. To assess in which part of the phase diagram non-local correlations become sizable in comparison to local ones, we calculate the local and nearest-neighbor (n.n.) self-energy in DiagMC and CDMFT in real space on the Matsubara axis. 
Fig.~\ref{fig:self_energy} displays the self-energy at the first Matsubara frequency {$\Sigma(i\omega_0\!=\!i\pi T)$} calculated by DiagMC (crosses) and CDMFT (squares) for two different temperatures (left panels: imaginary part, right panels: real part). 
The results of the calculations from both methods agree within 
error bars for both local and n.n. components. 

At high $T/t\!=\!0.40$ the correlations are mostly local and continuously increase from small to large $U$. However, there is an onset of non-locality already visible in the increase of the n.n. component  at the largest interactions shown. 
The non-local correlations remain very small at lower $T/t\!=\!0.10$ (close to the critical temperature of the MIT in CDMFT), until quite close to 
$U/t\!\approx\!9.25$ at which the MIT takes place in CDMFT. 
Hence, through most of the metallic regime except close to the MIT, the self-energy in this temperature range is local to a good approximation in this frustrated system.  

Upon entering the insulating regime, non-local correlations continuously increase.  
These non-local correlations signal increasing magnetic fluctuations. The onset of a magnetically ordered phase in DMFT (see App.~\ref{app:dmft}) underpins this interpretation. In the true solution of the system, of course, the Mermin-Wagner theorem \cite{Mermin1966, Hohenberg1967} prohibits magnetic ordering at finite temperature but the corresponding 
magnetic fluctuations are responsible for the increase of the (non-local) correlations. 
We note that the effect of magnetic fluctuations beyond DMFT using the dual fermion 
approximation has been investigated for this model in Refs.~\cite{Li2014,Laubach2015,Shaozhi2020} 
and that the implications of non-local effects for transport has been investigated in 
Ref.~\cite{Vranic2020}.


\section{Thermodynamics}
\label{sec:thermodynamics}

\begin{figure}[t]
    \centering
    \includegraphics[width=\columnwidth]{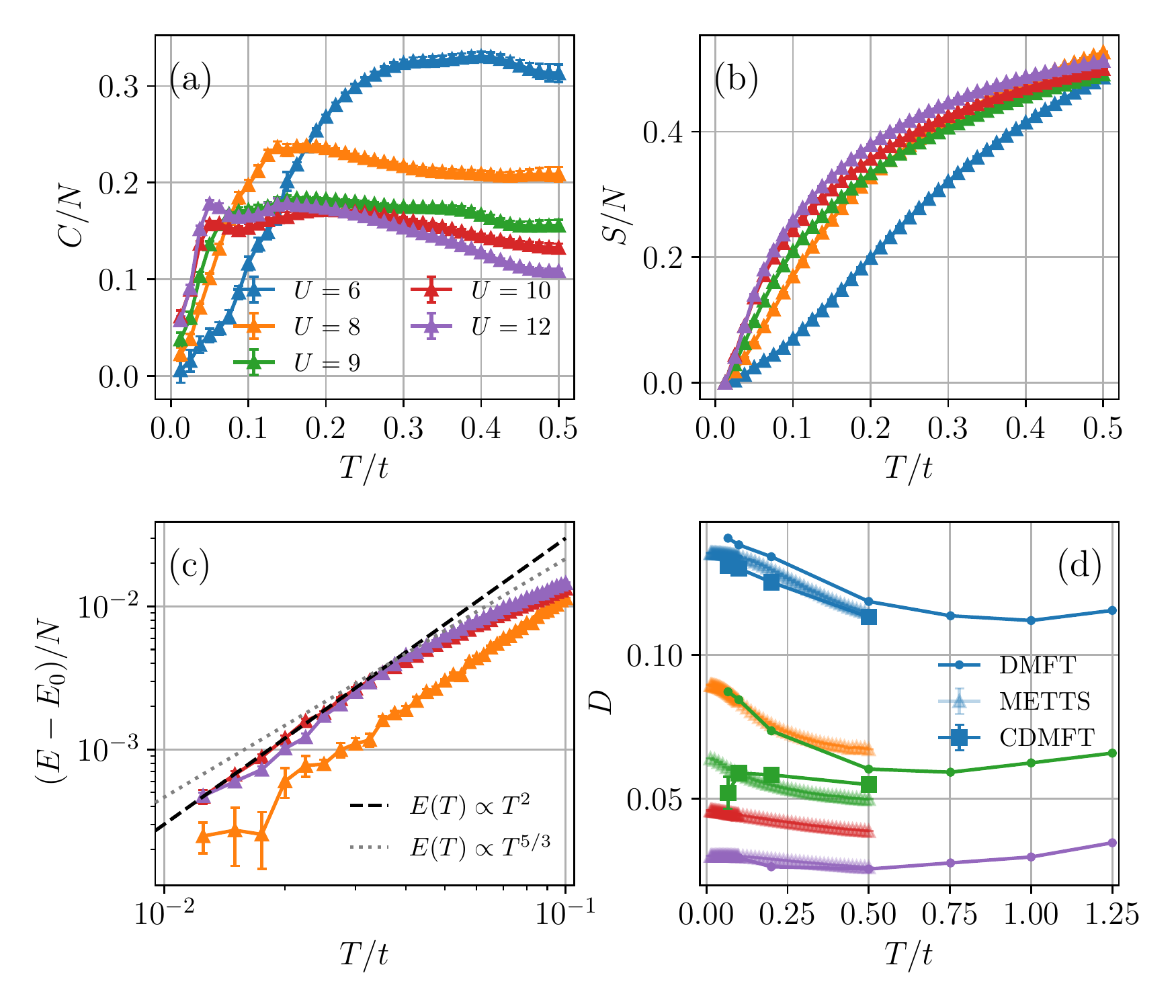}
    \caption{Thermodynamics of the $16\times 4$ YC4 cylinder from METTS 
    for different values of $U/t$. We employed a maximal bond dimension $D_{\max}=3000$ (a) Specific heat $C$. At $U/t=9,10,12$ we observe a broad continuum at higher temperatures with a small peak at $T/t\approx 0.05$.
    (b) Thermal entropy $S$. We observe an increase in entropy when increasing $U/t$ at low temperature. 
 (c) Internal energy $E$ as a function of temperature on a refined grid
    at lower temperatures. We observe regimes where approximately
    $E \propto T^2$, which implies $T$-linear behavior of the specific heat. The black and gray lines indicate $E \propto T^2$ and $E \propto T^{5/3}$ behavior.
    (d) Double occupancy $D$. At low temperatures, increasing $T/t$ decreases the 
    double occupancy. This order-by-disorder phenomenon is related to the 
    increase in thermal entropy with $U/t$ by the Maxwell relation 
    $\partial S / \partial U = - \partial D / \partial T$. We compare our data
    to results from (cluster) DMFT and find the CDMFT data closely matching the METTS
    results.}
    \label{fig:thermodynamics}
\end{figure}
We now turn to discussing the thermodynamic properties of the system for 
a range of interactions from $U/t=6$ to $U/t=12$.
Figure~\ref{fig:thermodynamics} displays the specific heat $C$, thermal 
entropy $S$, internal energy $E$, and double occupancy $D$ as a function of 
temperature. Results for the specific heat,
\begin{equation}
    C = \frac{\partial E}{\partial T},
\end{equation}
of the $16\times 4$ YC4 cylinder using METTS at various values of $U/t$ 
are shown in \cref{fig:thermodynamics}(a). At intermediate and large
interaction strengths 
$U/t=9, 10, 12$ the specific heat exhibits a large mostly featureless plateau
down to temperatures of $T/t \approx 0.1$. For $U/t=10,12$ a small peak
develops 
at $T/t\approx 0.05$ before the specific heat tends towards zero at $T=0$.

For $U/t=6, 8$ we find a $T$-linear behavior of $C$ at low temperature, 
consistent with a metallic phase with gapless excitations. 
We note that the low-$T$ slope for $U/t=8$ is approximately three times larger than that at $U/t=6$: this is 
qualitatively consistent with $Z_c$ (inset of Fig.~\ref{fig:mit_transition})
being approximately three times smaller. 
For a metal in which the self-energy can be approximated as local, the 
quasiparticle effective mass enhancement which controls the slope of $C$ is related to the quasiparticle weight $Z$ by $m^*/m=1/Z$. 
Indeed, as shown in the previous section, the non-local components of the self-energy are 
small through most of the metallic regime. 
Our findings for $Z_c$ and the slope of $C$ are thus consistent with 
quasiparticles developing a rather heavy mass as the insulator is approached. 
Although this is difficult to ascertain from our data, we find no evidence 
for a divergence of the effective 
mass when approaching the MIT however, consistent with the increasing non-locality of the self-energy 
in this regime (see Sec.~\ref{sec:selfenergy}). 
We also observe that for $U/t=6$, the specific heat appears to have another quasi-linear regime for $T/t\gtrsim 0.1$. 

At larger interactions strengths the low-temperature behavior of $C$ also
appears to be linear in $T$. However, given the few data points in this
regime, it is
difficult to discern this behavior from other scenarios. We note that the 
specific heat of the YC4 cylinder closely resembles the specific heat that has been
obtained on smaller clusters using the finite-temperature Lanczos method~\cite{Kokalj2013}.

We also investigate the thermodynamic entropy,
\begin{equation}
\label{eq:thermalentropy}
    S = \log (\mathcal{Z}) + \frac{E}{T} 
        = S_0 + \int_0^T \textrm{d}\Theta \frac{C(\Theta)}{\Theta},
\end{equation}
where $\mathcal{Z}$ denotes the partition
function and $E$ the internal energy. $S_0$ denotes a residual entropy at zero
temperature. The entropy is obtained by integrating the specific heat 
as in \cref{eq:thermalentropy} from $T=0$, where the
internal (ground state) energy is computed by DMRG and we assume $S_0=0$, 
i.e. we assume a unique ground state on the finite size cylinder. 

Our results from METTS on the YC4 cylinder are shown in
\cref{fig:thermodynamics}(b). 
Interestingly, we find that the entropy (at fixed $T$) increases rapidly with increasing
interaction strength from the metallic regime at $U/t=6$ to the insulating
regime beyond $U/t=9$. 
The increase in thermal entropy as a
function of interaction strength has previously also been observed using the
finite-temperature Lanczos method on smaller cluster geometries~\cite{Kokalj2013}. 
Naively, one would expect a decrease in entropy when the system is localizing. However,
we find the exact opposite behavior, i.e. $\partial S / \partial U > 0$. This
behavior is also reflected in the temperature dependence of the double
occupancy,
\begin{equation}
 D =  \frac{1}{N}\sum_{i=1}^N\braket{n_{i\uparrow}n_{i\downarrow}},
 \label{eq:doubleoccupancy}
\end{equation}
shown in \cref{fig:thermodynamics}(d). The Maxwell relation 
\begin{equation}
    \label{eq:maxwellrelation}
    \frac{\partial S}{\partial U}\bigg|_T = -\frac{\partial D}{\partial T}\bigg|_U
\end{equation}
relates the increase in entropy as function of $U$ to a decrease in 
double occupancy as a function of $T$. 
Indeed, as shown on \cref{fig:thermodynamics}(d), we observe a decrease of 
$D$ at low-$T$ upon heating, for all displayed values of $U/t$ up to a temperature
$T/t\approx 0.5$ at which the double occupancy has a minimum. 
On this figure, our METTS results are also compared to (C)DMFT which confirms our finding. 

This phenomenon is analogous to the Pomeranchuk effect in liquid Helium 3~\cite{Richardson1997}. 
When the entropy of the insulating state is larger than that of the metal, increasing temperature 
indeed leads to increased localisation since this yields a gain in free-energy. 
To the best of our knowledge, this behavior was first predicted for the Hubbard model 
on the basis of DMFT studies~\cite{Georges1992_Mott} 
(see also \cite{Georges1993,Rozenberg1994,Dare2007,Georges2011,Li2014}). 
It has been proposed~\cite{Werner2005} and experimentally realized~\cite{Taie2012} 
as a cooling scheme in the context of cold atomic gases in optical lattices. 
Interestingly, it has also been recently observed in 
magic angle graphene~\cite{Rozen2020}. 
In the present model, the large entropy of the insulating phase due to frustration and 
competing orders 
is responsible for this effect. 

An important aspect of the thermodynamics is the behavior of the specific heat at low temperatures. In particular, the behavior of the specific
heat allows us to distinguish between gapped and gapless phases. Here, we investigate the internal energy $E$ instead of the specific heat
$C=\partial E / \partial T$, since we measure the energy directly, whereas
the specific heat is obtained from a numerical derivative of the energy. 
Our results for $U/t=8, 10, 12$ are shown in \cref{fig:thermodynamics}(c).
An analysis of the convergence as a function of bond dimension for $U/t=10$
is shown in the appendix in \cref{fig:energyconvergence}. In the case of 
$U/t=8$, we find that the internal energy $E$ approximately behaves as $E \propto T^2$ for temperatures $0.02 \leq T/t \leq 0.1$, which is indicated by the black dashed lines. This translates to a $T$-linear behavior of the specific heat upon differentiation. Similarly, for $U/t=10$ and $U/t=12$ the energy is well-descibed by $E \propto T^2$ behavior for $0.0125 \leq T/t \leq 0.4$. We notice, that the temperatures in this regime coincide with region below the peak in the specific heat in \cref{fig:thermodynamics}(a). 
For comparison, we also show
a scaling $E \propto T^{5/3}$ (i.e. $C\propto T^{2/3}$) as a gray dotted line, which is the expected 
behavior of a spinon Fermi surface
state~\cite{Motrunich2005,Sheng2009,Block2011}. As can be seen in \cref{fig:thermodynamics}(c) our data is in closer agreement to a $E \propto T^2$ than the $E \propto T^{5/3}$ scenario. The $T$-linear
behavior we observe would indicate a gapless state. However, we would like to point out that the lowest temperature attained in these simulations is $T/t=0.0125$. Hence, our data does not rule out an activated behavior from a gap smaller than this temperature.  
We would like to point out, that these results are in agreement with experimental measurements of the specific heat for  the triangular lattice compound \ch{$\kappa$-(ET)2Cu2(CN)3}~\cite{Yamashita2008}.

\begin{figure}[t]
    \centering
    \includegraphics[width=\columnwidth]{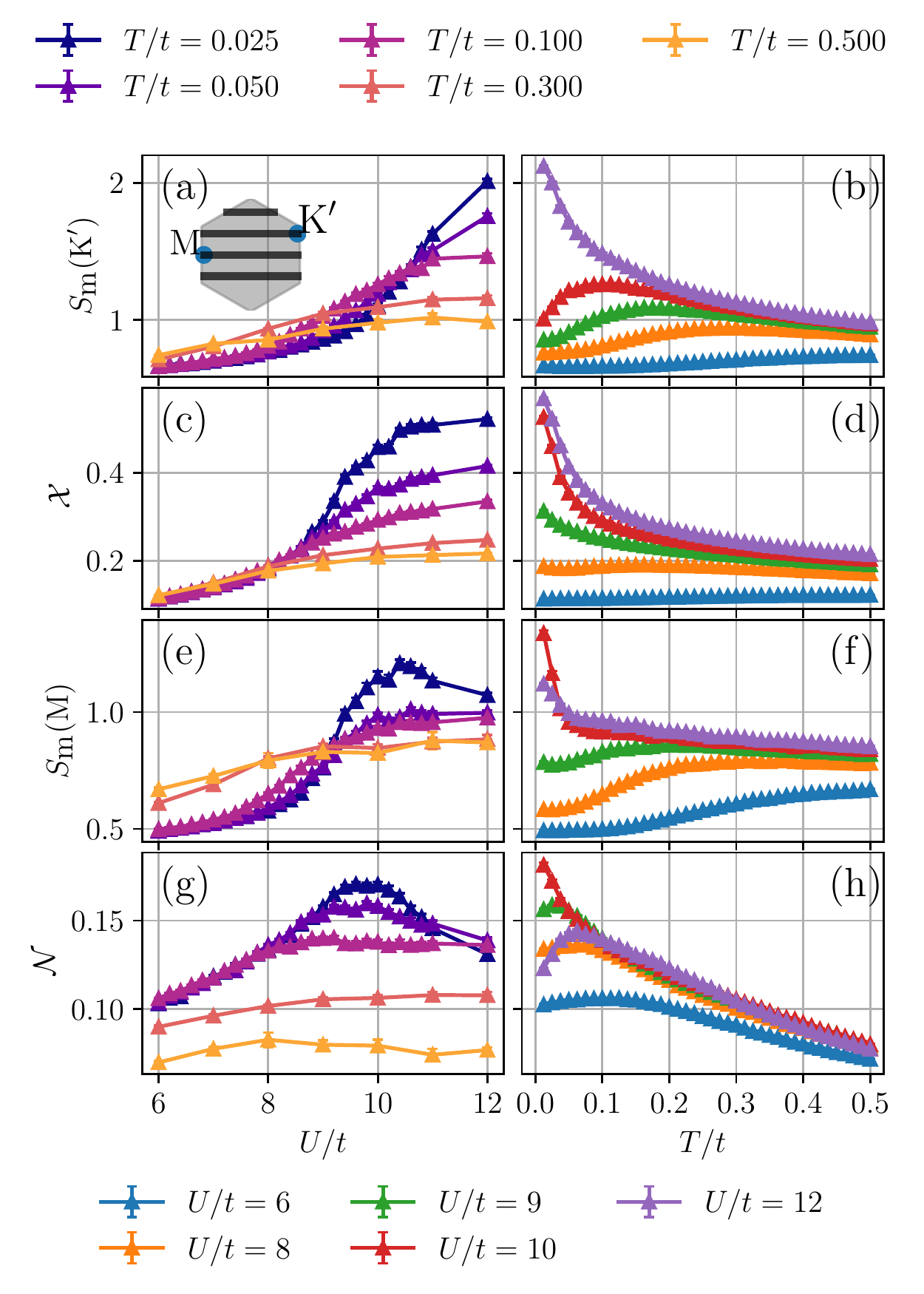}
    \caption{Magnetic ordering as a function of interaction strength $U/t$ (left)
    and temperature $T/t$ (right) on the $16 \times 4$ YC4 cylinder from METTS.
    Simulations have been performed with a maximal bond dimension 
    $D_{\max}=3000$. (a,b) Magnetic structure factor $S_{\textrm{m}}(\textrm{K}^\prime)$ 
    indicating $120^\circ$ N\'eel order. The inset in (a) shows the momenta
    resolved by the YC4 cylinder, and the position of the ordering vectors
    $\textrm{K}^\prime$ and $\textrm{M}$
    (c,d) chiral susceptibility $\mathcal{X}$ as defined in
    \cref{eq:chiralsusceptibility}. Chiral correlations build up in both
    the intermediate as well as the strongly coupled regime at lower temperature.
    (e,f) Magnetic structure factor $S_{\textrm{m}}(\textrm{M})$ indicative of
    collinear stripy antiferromagnetic order. (g,h) Nematic spin correlation 
    $\mathcal{N}$, as defined in \cref{eq:nematic}, is pronounced only in the 
    intermediate coupling regime.
    }
    \label{fig:magnetism}
\end{figure}

\section{Magnetism}
\label{sec:magnetism}

\begin{figure*}[t]
    \centering
    \includegraphics[width=\textwidth]{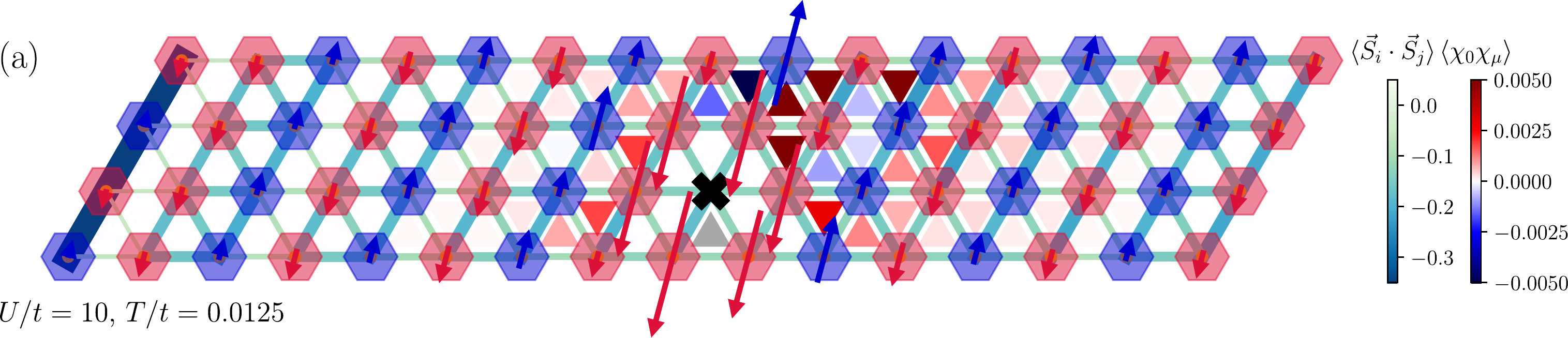} 
    \includegraphics[width=\textwidth]{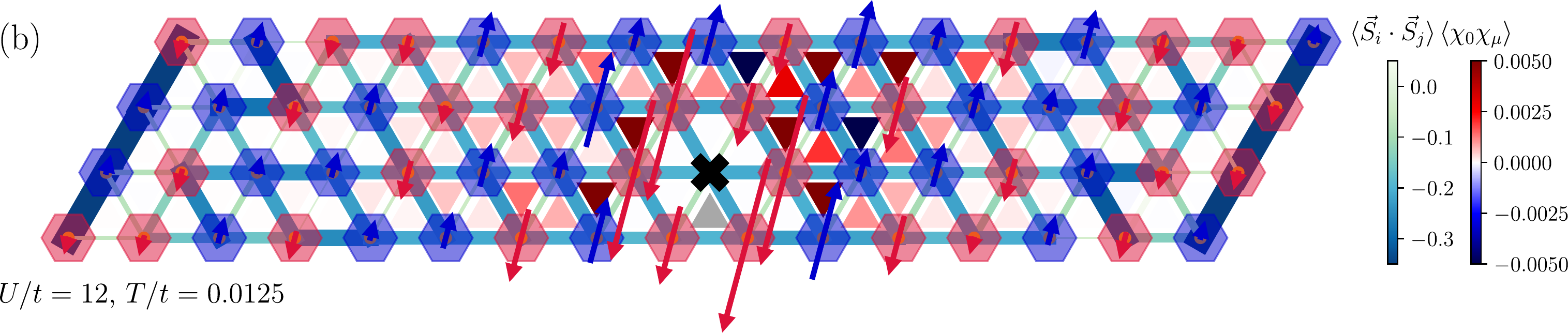}
    \caption{Snapshots of METTS states $\ket{\psi_i}$ at temperature $T/t=0.0125$
    on the $16\times4$ YC4 cylinder in the intermediate regime at $U/t=10$ (a)
    and in the strong coupling regime at $U/t=12$ (b). The 
    length of the arrows is proportional to the spin correlation 
    $\braket{\vec{S}_0 \cdot \vec{S}_i}$, where
    the black cross marks the reference site.
    Hexagons and arrows which are blue indicate positive and 
    red indicate negative spin correlations.
    The color of the triangles indicates the magnitude of the chiral correlation $\braket{\chi_0 \chi_\mu}$, where the reference triangle is indicated in gray just below the reference site.
    Nearest-neighbor spin correlations $\braket{\vec{S}_i \cdot \vec{S}_j}$ are 
    indicated as the width and color of the bonds. We observe collinear stripy
    spin correlations at this temperature in the intermediate regime at $U/t=10$.}
    \label{fig:snapshots}
\end{figure*}

We study the magnetic properties as a function of temperature and interaction 
strength of the system on the YC4 cylinder using the METTS algorithm. 
Results on magnetic ordering from (dynamical) mean-field theory can 
be found in Appendix~\ref{app:dmft}. To distinguish different kinds of
orderings, we investigate the magnetic structure factor,
\begin{equation}
  \label{eq:magnstructure}
  S_{\textrm{m}}(\bm{k}) = \frac{1}{N}\sum_{l,m=1}^N
  \text{e}^{i\bm{k} \cdot (\bm{r}_l - \bm{r}_m)}
  \braket{\vec{S}_l\cdot\vec{S}_m}.
\end{equation}
The momenta $\bm{k}$ resolved by the YC4 cylindrical geometry are shown in 
the inset of \cref{fig:magnetism}(a). Magnetic $120^\circ$ N\'eel order can be detected 
by observing a peak in the structure factor at the $\textrm{K}$ point 
in the Brillouin zone. On the YC4 cylinder, 
the $\textrm{K}$ point is not exactly resolved, which is why we resort to the 
closest point $\textrm{K}^\prime$, shown in \cref{fig:magnetism}(a), to indicate
$120^\circ$ N\'eel order. A peak at the $\textrm{M}$-point can indicate the following two 
kinds of magnetic correlations:
\begin{enumerate}
    \item[(i)] A collinear `stripy' antiferromagnetic
ordering is characterized by breaking both spin and discrete $C_6$ lattice
rotation symmetry. This kind of ordering is 
characterized by the spins being aligned ferromagnetically along one
direction of the triangular lattice and antiferromagnetically along the other two. 
Note that we use here the term `stripy' in relation to spin degrees of freedom - we find no 
indication of a charge stripe density modulation. 
  \item[(ii)] Non-coplanar tetrahedral order, on the other hand, is 
formed when spins in a $2 \times 2$ unit cell align in a way that they point
towards the corners of a regular tetrahedron~\cite{Messio2011,Wietek2017}. 
\end{enumerate}
Several recent DMRG studies~\cite{Szasz2020,Szasz2021,Chen2021} have
demonstrated that the triangular Hubbard model in the intermediate regime is
susceptible to time-reversal symmetry breaking, which is indicated by
a non-zero expectation value of the scalar chirality operator in the 
thermodynamic limit. To study such a scenario, we 
compute the chiral susceptibility
 \begin{equation}
 \label{eq:chiralsusceptibility}
    \mathcal{X} = \frac{1}{N}\sum_{\mu,\nu \in \triangle} \braket{\chi_\mu \chi_\nu},
 \end{equation}
where the scalar chirality operator on a triangle \mbox{$\mu=(l,m,n)$} is given by
\begin{equation}
    \chi_\mu = \vec{S}_l \cdot (\vec{S}_m \times \vec{S}_n).
\end{equation}
The sum in \cref{eq:chiralsusceptibility} extends over all pairs of 
elementary triangles. In the case of spontaneous time reversal breaking, we
expect long range chiral correlations indicated by a large value of
the chiral susceptibility $\mathcal{X}$. Since our simulations are 
working with real-valued wave functions, the expectation values of the 
scalar chirality operators, $\braket{\chi_\mu}$, are exactly zero.

Moreover, we noticed a particular feature in our data on the YC4 cylinder, 
where prominent nearest-neighbor antiferromagnetic correlations switch between
the different directions of the triangular lattice. To quantify this observation
we introduce the nematic spin correlation,
\begin{equation}
    \label{eq:nematic}
    \mathcal{N} = -\frac{1}{N}\sum_{\{ i,j \}^\prime } \braket{\vec{S_i}\cdot \vec{S_j}},
\end{equation}
where the sum extends only over nearest-neighbor pairs $\{ i, j\}^\prime$ along
the short direction of the cylinder (i.e. the direction pointing ``north-east'' in \cref{fig:snapshots}). 
The behavior of the above quantities as both a function of $U/t$ and
temperature $T/t$ is shown in \cref{fig:magnetism}. 
We observe three distinct regimes as a function of $U/t$. 
For $U/t \lesssim 8.5$ we do not observe any  dominant magnetic features. 
As we have previously found in \cref{fig:mit_transition} and \cref{fig:dofk},
this region corresponds to the metallic regime.

The intermediate regime ranging from $8.5 \lesssim U/t \lesssim 10.5$ exhibits interesting and peculiar behavior in all
observables. 
With decreasing temperature,
magnetic correlations grow as signaled by the structure factor at both the $K'$ and $M$ 
points \cref{fig:magnetism}(b,f).
But below the temperature $T/t=0.1$  for $U/t=10$, the $K'$-point
structure factor begins to decrease, while at the $M$-point it increases sharply beginning around $T/t=0.05$. The chiral correlations
in \cref{fig:magnetism}(d) also increase below this scale, and the
specific heat simultaneously develops a small maximum then rapidly decreases as shown in \cref{fig:thermodynamics}(a). 
The development of low-T chiral correlations is 
consistent with previous DMRG results~\cite{Szasz2020,Chen2021}, 
which proposed that at $T=0$ the system spontaneously breaks time-reversal
symmetry and forms a chiral spin liquids. However, we also observe that 
the chiral correlations similarly build up beyond $U/t\gtrsim 10.5$.

As already pointed out before, in principle a peak at the $\textrm{M}$ point in the intermediate regime could 
also indicate non-coplanar tetrahedral magnetic order~\cite{Messio2011}. 
However, by inspecting real space spin correlations we clearly observe the
formation of stripy antiferromagnetism. Using METTS we can investigate
``snapshots'' of the system at a given temperature~\cite{Wietek2020}. Briefly
summarized, the METTS method decomposes the thermal density matrix into a
sum over rank-1 density matrices corresponding to pure
states~\cite{White2009,Stoudenmire2010},
\begin{equation}
    \frac{1}{\mathcal{Z}}\textrm{e}^{-\beta H} = \sum_i p_i \ket{\psi_i}\bra{\psi_i},
\end{equation}
where $p_i\geq 0$ are real non-negative probabilities, $\ket{\psi_i}$ are 
the so-called METTS wave functions, and $\mathcal{Z}$ denotes the partition
function. The pure states $\ket{\psi_i}$ are sampled 
with probability $p_i$. We show the properties of a typical
METTS wave function sampled in our simulations at $T/t=0.0125$ and $U/t=10$ in
\cref{fig:snapshots}(a).
The stripy spin correlations are clearly pronounced for this METTS 
state. We also observe sizeable chiral correlations $\braket{\chi_0 \chi_\mu}$
which are indicated by the color of the inner triangles. When comparing the
snapshot at $U/t=10$ in \cref{fig:snapshots}(a) to 
the snapshot at a larger $U/t=12$ in \cref{fig:snapshots}(b) we observe that the nearest-neighbor 
spin correlations are more strongly pronounced along the short direction of the
cylinder at $U/t=10$, whereas for $U/t=12$ the spin correlations on the other 
two directions are enhanced.

This motivates the definition of the nematic spin correlation
$\mathcal{N}$ in \cref{eq:nematic}. In \cref{fig:magnetism}(g,h)
we observe that $\mathcal{N}$ is clearly pronounced only in the intermediate
regime. We would like to point out that the increased magnetic correlations 
align with the stripy spin patterns, as can be seen in \cref{fig:snapshots}(a).

One last notable aspect of the intermediate regime   
we find is that increasing the temperature from $T=0$ not only suppresses double occupancy, as shown previously in \cref{fig:thermodynamics}, but also increases the $120^\circ$ N\'eel order correlations. Therefore, the effect of increasing 
temperature is similar to the effect of further increasing the coupling 
strength $U/t$, which also both localizes the system and favors $120^\circ$ N\'eel order for $U/t \geq 10.5$.

The strong coupling regime  $U/t \gtrsim 10.5$ is most prominently characterized by 
the increase in the magnetic structure factor $S_{\textrm{m}}(\textrm{K}^\prime)$
at lower temperatures, shown in \cref{fig:magnetism}(a,b). This is indicative of
$120^\circ$ N\'eel order in the ground state. We observe strong antiferromagnetic
correlations for $U/t=12$ setting in at a temperature below $T/t=0.05$, which
again coincides with the small maximum in the specific heat observed in
\cref{fig:thermodynamics}(a). We find that the behavior of the chiral 
correlations is similar to the intermediate coupling regime. In particular,
$\mathcal{X}$ as shown in \cref{fig:thermodynamics}(d) is rather comparable 
between $U/t=10$ and $U/t=12$.

\section{Discussion}
\label{sec:discussion}
The physics of the triangular lattice Hubbard model at half-filling is coarsely
organized in three different regimes as a function of the coupling strength,
$U/t$: a metallic regime  
is followed by an intriguing insulating regime at intermediate coupling regime 
whose nature is currently hotly debated.
At large interaction strength the system enters a magnetic insulating regime, 
where coplanar $120^\circ$ N\'{e}el order is stabilized in the ground state.
Evidence for the existence of an intermediate non-magnetic insulating regime is
ample in the literature~\cite{Morita2002,Kyung2006,Sahebsara2008,Li2014,Laubach2015,Misumi2017,Tocchio2008,Yoshioka2009,Yang2010,Szasz2020,Chen2021} 
and clearly confirmed by several of our findings using multiple numerical
methods. 

We firmly establish the order-by-disorder effect at intermediate coupling
$U/t$, where increasing temperature paradoxically leads to increased localization, 
as apparent in the double occupancy shown in \cref{fig:thermodynamics}(d). 
As discussed in Sec.~\ref{sec:thermodynamics}, this effect is similar to the Pomeranchuk effect~\cite{Richardson1997} observed when liquid Helium 3 solidifies upon heating, and 
previously found to occur for the Hubbard model 
in DMFT studies~\cite{Georges1992_Mott,Georges1993,Rozenberg1994,Werner2005,Dare2007}. 
The decrease in double occupancy upon heating is confirmed by both our METTS and cluster DMFT results, 
where we found good
quantitative agreement between these two very different numerical techniques. 
This observation suggests that localized excitations carry a large thermal entropy 
at low temperatures. This is consistent with the Maxwell relation 
\cref{eq:maxwellrelation} relating decreasing double occupancy in temperature to an
increasing entropy with interaction strength $U$, which we confirm by computing the thermal entropy 
from METTS in \cref{fig:thermodynamics}(b), as also 
previously observed 
using the finite-temperature Lanczos method on smaller cluster geometries~\cite{Kokalj2013}. 
Upon increasing the temperature we observe increased $120^\circ$ N\'{e}el correlations for $U/t=10$ in \cref{fig:magnetism}(b). Hence, both
increasing temperature as well as increasing $U/t$, increase the systems
tendency to localize and, therefore, form $120^\circ$ order. The formation
of $120^\circ$ magnetic order can thus be seen as an analogy of liquid Helium 3 solidifying upon increasing temperature.

This order-by-disorder effect naturally gives rise to the question about the nature 
of the proliferating excitations causing the localization at finite temperature. 
Let us first discuss the intermediate coupling regime at $U/t=10$.  As we have 
shown in \cref{fig:magnetism}, both the chiral correlations as well as the 
magnetic structure factor at the $\text{M}$ point develop a maximum towards $T=0$
at $U/t=10$ on the $16\times 4$ YC4 cylinder. Interestingly, finite temperature 
does not simply melt this ordering. Instead, we observe a maximum in
$S_{\textrm{m}}(\textrm{K}^\prime)$ at $T/t=0.1$ indicating increased $120^\circ$
spin correlations. It is interesting to note, that increasing temperature appears
to have the same effect as increasing interaction strength, which also favors
 $120^\circ$ order. 
 
Let us now turn to discussing the orders which may develop in the intermediate regime $8.5 \lesssim U/t \lesssim 10.5$.
With decreasing temperature, we find increased chiral correlations 
as well as spin correlations at the $\textrm{M}$ point. While the onset of chiral
correlations would be consistent with spontaneous time-reversal and parity symmetry
breaking, as expected for a ground-state chiral spin liquid (CSL)~\cite{Szasz2020,Szasz2021,Chen2021},
the peak in $S_{\textrm{m}}(\textrm{M})$ is not
necessarily related to the formation of a CSL. While a peak at the $\textrm{M}$ 
point could in principle be indicative of non-coplanar tetrahedral
order~\cite{Messio2011}, we find that in the present geometry this peak is 
related to the formation of nematic, stripy antiferromagnetic correlations.
This finding is backed up by a recent ground state DMRG study~\cite{Chen2021},
which analogously found a peak in $S_{\textrm{m}}(\textrm{M})$ in the
intermediate regime, although the authors found these correlations 
to be only short-ranged at $T=0$.

The occurrence of stripy spin correlations is remarkable given that most
known instances of chiral spin liquids are stabilized in close proximity to
non-coplanar
magnetically ordered states~\cite{Messio2012,He2014,Gong2014,Wietek2015,Wietek2017,Gong2017}. 
The melting of non-coplanar magnetic ordering has even been suggested as
a guiding principle to understand the formation of CSLs~\cite{Hickey2017}. 
This seems to be rather different in the present case, where nematic collinear 
correlations and chiral correlations both develop as the temperature is decreased. 
In this context, the variational study of
the triangular lattice Heisenberg model with an additional ring-exchange term performed in Ref.~\cite{Grover2010} is
particularly interesting. The model with ring exchange can be thought of an approximate low-energy 
effective Hamiltonian for the intermediate coupling regime~\cite{Motrunich2005,Yang2010}.
The authors compared variational energies of several Gutzwiller projected 
ansatz wave functions, including an ansatz for a gapped chiral spin liquid 
and a gapless nematic spin liquid, breaking
rotational symmetry. While both of these wave functions have been shown to have
a comparable, competitive energy, the gapless nematic state had the
lower variational energy for this particular model. A more recent variational Monte Carlo 
study has similarly suggested the stabilization of a gapless nematic
spin liquid in the context of the half-filled triangular Hubbard model, albeit 
upon adding further second nearest-neighbor interaction~\cite{Tocchio2020}.
Remarkably,  our finite-temperature METTS simulations now reveal exactly this
competition between chiral and stripy spin correlations at finite, but low
temperatures. While the recent DMRG studies~\cite{Szasz2020,Chen2021} provided
strong evidence, that ultimately at $T=0$ a CSL is formed on the investigated
geometries, we now propose that stripy spin correlations become relevant immediately at finite
temperatures. We would like to point out, that an estimate for the gap 
of the CSL has been stated in Ref.~\cite{Szasz2020} by computing the domain wall tension to be $\Delta \approx 0.0065t$,
which is below the lowest temperature $T/t = 0.0125$ we have been able to simulate
using METTS on the $16\times 4$ cylinder. It is worth pointing out that the effect 
of finite-temperatures on the system is non-trivial, as can be seen by the
increase in $S_{\textrm{m}}(\textrm{K}^\prime)$ when increasing temperatures in
\cref{fig:magnetism}(a), or the decrease of the double occupancy in
\cref{fig:thermodynamics}(d). 

This raises the question about whether a (rotationally-symmetric)
perturbation could stabilize nematic stripy (quasi-)order at $T=0$. 
In particular, it would be interesting to find out whether indeed a nematic
gapless spin liquid with algebraic spin correlations can be realized and to study
its transition to the CSL. However, such a state will likely have larger
quantum entanglement than the gapped CSL, rendering accurate DMRG 
computations more difficult. 

We expect the balance between chiral or nematic spin correlations to be strongly 
dependent on the finite size geometry. As the DMRG studies on the YC3-6 and XC4
cylinders have shown~\cite{Szasz2020,Szasz2021,Chen2021}, the precise nature of the
ground state still has rather strong dependence on the exact shape of the 
cylindrical geometry. Therefore, a detailed comparison of our finite-temperature
METTS data for different geometries is therefore highly desirable. 

We have performed an in-depth comparison between different geometries in \cref{sec:geometries}. There, we discuss results on YC4 cylinders of varying
length and also results on the YC3 and XC4 cylinders~\cite{Szasz2020}. We find
that our results only weakly depend on the length of the YC4 cylinders. The 
specific heat of the YC4 cylinder closely resembles the specific heat obtained
on the YC3 and XC4 cylinders. In particular, we observe that the maxima for 
different values of $U/t$ develop at comparable temperatures across all different
geometries. Similarly, the Pomeranchuk effect of decreasing double occupancy 
as a function of temperature is clearly observed on all geometries. The correlations
at low temperatures in the intermediate coupling regime differ, however. 
While both the YC4 and YC3 geometries exhibit pronounced stripy antiferromagnetic
correlations, this is not observed on the XC4 cylinder. Also, the chiral
susceptibility only smoothly increases as a function of $U/t$ on the YC3 and XC4
cylinders. This is in contrast to the YC4 cylinder, where we observe the onset of
chiral correlations at smaller $U/t$ than the onset of $120^\circ$ magnetic 
correlations.

At this point, we would
like to comment that finite-size effects are expected to become less severe at higher
temperatures, since correlation lengths typically decrease. It remains to be seen
down to which temperature scale the finite-size cylinders can fully capture the two-dimensional limit. 

Let us now focus on the strong coupling regime at $U/t = 12$. 
In the limit $U/t\rightarrow \infty$ the effective spin degrees of freedom are described by the
antiferromagnetic Heisenberg model, whose $120^\circ$ N\'{e}el ordered ground
state features spin-wave excitations. However, besides spin-wave excitations
several authors have found a different kind of excitation being relevant in this
case. Series expansions found anomalous behavior of the magnon spectrum,
which exhibits a minimum beyond the description of linear spin-wave
theory~\cite{Zheng2006a,Zheng2006b,Starykh2006}. This minimum is ascribed
to the presence of a different kind of excitations, reminiscent of the
roton excitations of \ch{^4He} forming a minimum in the quasiparticle
dispersion~\cite{Landau1941,Feynman1957}.
Therefore, the excitations of the triangular lattice Heisenberg antiferromagnet have
often been referred to in literature as ``rotonlike'' excitations (RLE). It has been
argued that these excitations contribute significantly to the thermal entropy
down to temperatures $T \approx 0.1 J$~\cite{Zheng2006b}, where $J$ denotes the
antiferromagnetic coupling constant. More thoroughly, the presence of two
different kinds of excitations in the $S=1/2$ triangular Heisenberg antiferromagnet 
has been recently using the exponential tensor renormalization group (XTRG)
method~\cite{Chen2018XTRG,Chen2019}, which similar to METTS, allows for unbiased numerical
simulations at finite-temperature on cylindrical geometries. The authors indeed
establish two temperature scales corresponding to the magnon and the RLEs. The RLEs
are shown to manifest themselves in an increase of the nearest-neighbor chiral
correlations as well as a maximum in the magnetic structure factor
$S_{\textrm{m}}(\textrm{M})$~\cite{Chen2019}.
The existence of the RLEs is demonstrated to be robust on a wide variety of 
cylinder geometries, including cylinders of circumference $W=6$. %

In our simulations at $U/t=12$ we analogously find a maximum in
$S_{\textrm{m}}(\textrm{M})$ in \cref{fig:magnetism}(e) indicating the RLEs. 
Also an increase of chiral correlations in \cref{fig:magnetism}(d) at temperatures
below $T/t\approx 0.1$ is observed. The $120^\circ$ magnetic correlations at lower
temperatures are signalled by a peak in $S_{\textrm{m}}(\textrm{K}^\prime)$. This
clearly resembles the situation encountered in the Heisenberg model as in
Ref.~\cite{Chen2019}. Interestingly, we found the specific heat in
\cref{fig:thermodynamics} to be rather similar for both the intermediate and strong
coupling regime. It appears that tuning the interaction strength from $U/t=10$ to
$U/t=12$ interchanges the role of stripy correlations with $120^\circ$ correlations.

We would like to elaborate on the behavior of the chiral susceptibility $\mathcal{X}$. If time-reversal symmetry is indeed broken at low-temperature
in the intermediate regime, we would expect the chiral correlations to diverge towards $T=0$. However, we think that the temperatures ($T\geq0.0125t$) we simulated are above the transition temperature estimated by Ref.~\cite{Szasz2020}, $\Delta \approx 0.0065t$. Furthermore,
it has also been argued that the stabilization of a CSL at $T=0$ can
depend on the cylinder length studied in DMRG~\cite{Chen2021}, where a
CSL at $T=0$ has only been found in cylinders of length $L=64$, but not
in shorter cylinders. Nevertheless, we find pronounced chiral correlations
in the intermediate coupling regime. For the strong coupling regime, 
pronounced chiral correlations at finite-temperature have been
found in a previous study of the Heisenberg model using
XTRG~\cite{Chen2019}. There, the correlations have been attributed to the
rotonlike excitations of the triangular lattice Heisenberg antiferromagnet.
Here, we find that the chiral correlations already build up in
the intermediate regime and remain sizeable in the strong coupling regime.
This strongly suggests that the rotonlike excitations are relevant
excitations in the intermediate coupling regime.

Finally, an outstanding question is the occurrence of superconductivity in the present model. 
While general arguments suggest that the metallic phase studied here hosts a low-temperature superconducting instability at weak coupling (see e.g. \cite{Raghu2010}), the possible occurrence 
of an unconventional superconducting phase near the metal-insulator phase 
boundary~\cite{Kyung2006} or upon doping the insulating phase~\cite{Zhu2020} are
intriguing questions for future computational studies.

\section*{Acknowledgements}
We are indebted to Nils Wentzell, Steven White, Michael Zaletel, and Sabine Andergassen for insightful discussions and support. We thank Elio K{\"o}nig for valuable comments on the manuscript. We thank the computer service facility of the MPI-FKF, and the Scientific Computing Core of the Flatiron Institute for their help. 
METTS results were obtained using the ITensor Library (C++ version) \cite{itensor} 
and CDMFT computations used the TRIQS library\cite{TRIQS}. 
This work was granted access to the HPC resources of TGCC and IDRIS under the allocations
A0090510609 attributed by GENCI (Grand Equipement National de Calcul Intensif). The authors gratefully acknowledge use of the computational resources of the Max Planck Computing and Data Facility. The present work was supported by the Austrian Science Fund (FWF) through the Erwin-Schr\"odinger Fellowship J 4266 - ``{\sl Superconductivity in the vicinity of Mott insulators}'' (SuMo, T.S.). It also has been supported by the Simons Foundation within the Many Electron Collaboration framework. 
A.G. also acknowledges the support of the European Research Council (ERC-QMAC-319286). 
The Flatiron Institute is a division of the Simons Foundation. 

\appendix
\section*{Appendix}

\section{Convergence of METTS simulations}
\label{app:metts}
\begin{figure}[t!]
    \centering
    \includegraphics[width=\columnwidth]{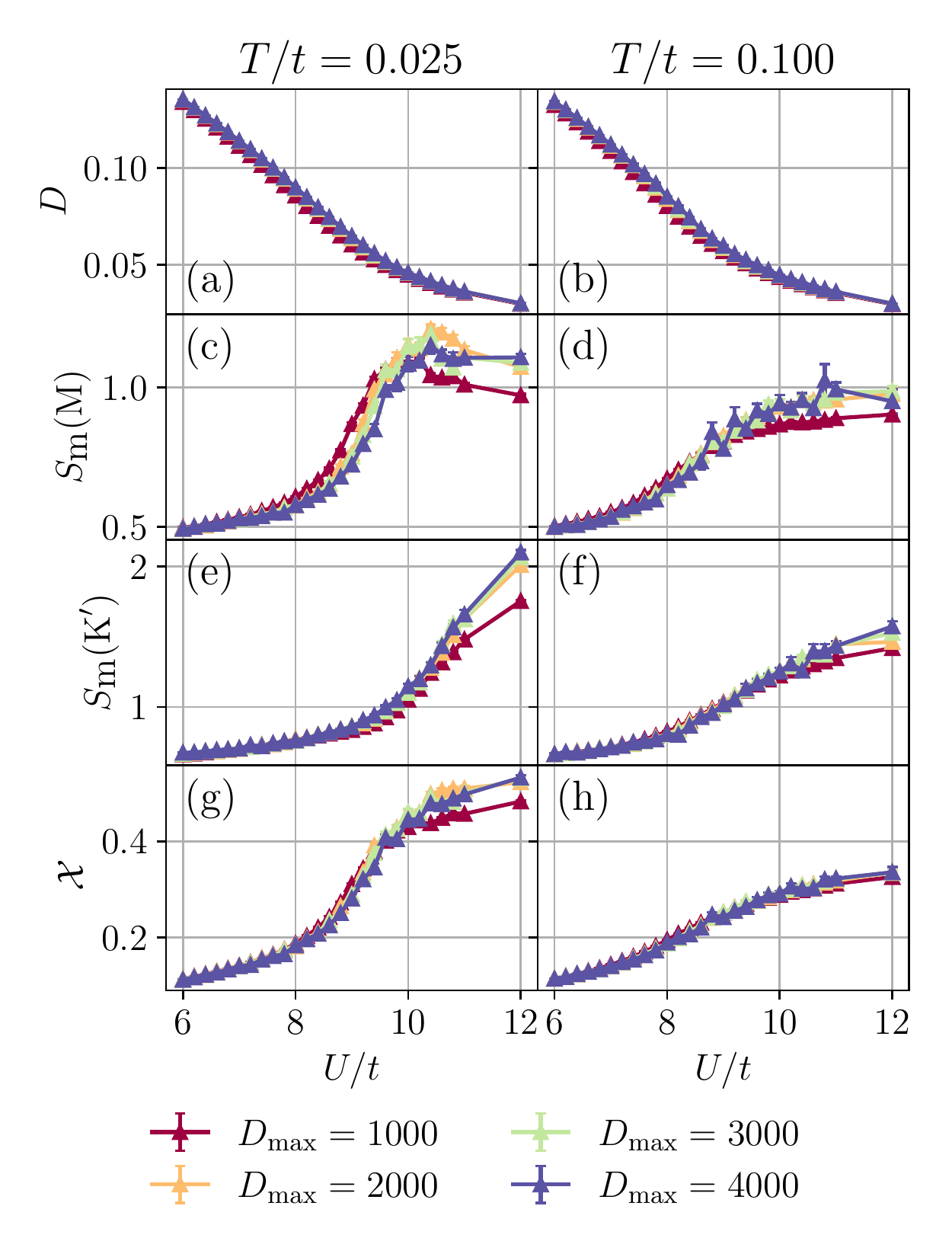}
    \caption{Convergence of METTS results on the $16\times 4$ YC4 cylinder
    as a function of maximal bond dimension $D_{\max}$. We compare results 
    from simulations performed with $D_{\max}=1000, 2000, 3000, 4000$. Results
    agree for all quantities within errorbars for $D_{\max}=2000, 3000, 4000$,
    whereas results at $D_{\max}=1000$ deviate slightly. Comparisons are
    performed at $T/t=0.025$ 
    (left) and $T/t=0.1$ (right). We show the double occupancy $D$ (a,b), 
    the magnetic structure factors evaluated at $\textrm{M}$ (c,d) and
    $\textrm{K}^\prime$(e,f), and the chiral susceptibility $\mathcal{X}$ (g,h).}
    \label{fig:convergence}
\end{figure}
We employ the METTS algorithm as described in Ref.~\cite{Wietek2020}. Thermal expectation values of an operator $\mathcal{O}$ are evaluated as,
\begin{equation}
  \braket{ \mathcal{O} } = \overline{\braket{\psi_i | \mathcal{O} | \psi_i}},
\end{equation}
where the minimally-entangled typical thermal states, 
\begin{equation}
  \label{eq:mettsstateimagtime}
  \ket{\psi_i} = \text{e}^{-\beta H / 2} \ket{\sigma_i},
\end{equation}
are imaginary-time evolved product states $\ket{\sigma_i}$. Here,
$\overline{\cdots}$ denotes statistical averaging over a series of subsequent
METTS. As such, the METTS algorithm is subject to statistical sampling uncertainty,
which can be reduced by computing more samples and whose size
 can be estimated using standard time series analysis. 
The imaginary-time evolution is performed by using the time-dependent
variational principle (TDVP) algorithm for matrix product
states~\cite{Haegeman2011,Haegeman2016,Paeckel2019}. In
Ref.~\cite{Wietek2020}, some of us showed that the maximal bond dimension 
$d$ of the matrix product state representation of the METTS serves as 
a control parameter to achieve accurate and controlled computations on finite
size cylinders. Here, we performed extensive comparisons between simulations
at different bond dimensions. Results on the $16 \times 4$ YC4 cylinder at 
temperatures $T/t=0.025$ and $T/t=0.1$ as a function of $U/t$ are shown in
\cref{fig:convergence}. Simulations have been performed up to a maximal bond
dimension of $D_{\max}=4000$. We find that all quantities of interest are converged
within errorbars already at $D_{\max}=2000$. 

\begin{figure}[t!]
    \centering
    \includegraphics[width=\columnwidth]{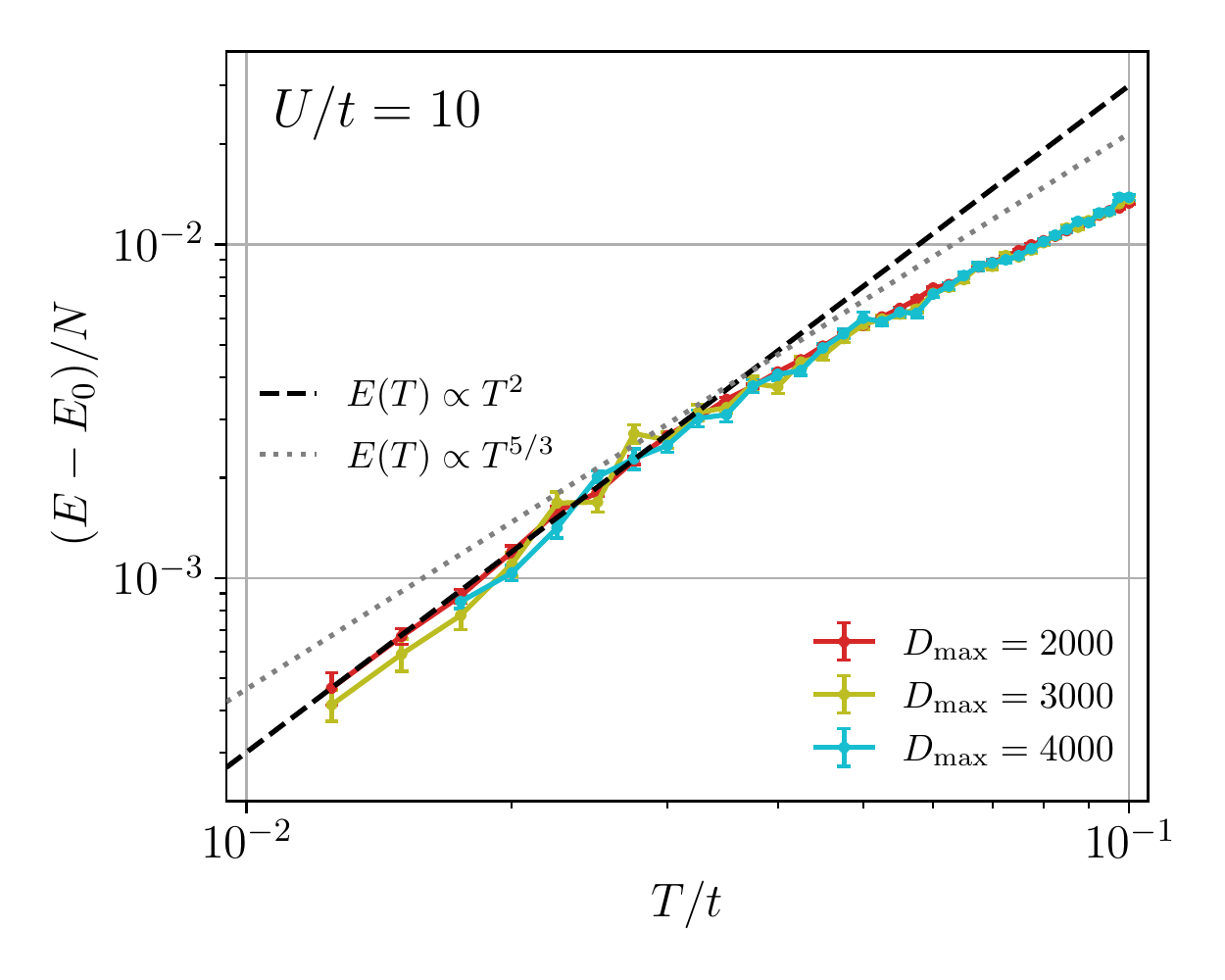}
    \caption{The internal energy $E$ as a function of $T/t$
    at $U/t = 10$ on the $16\times 4$ YC4 cylinder
    depending on the maximal bond dimension $D_{\max}$. We compare results 
    from simulations performed with $D_{\max}=2000, 3000, 4000$ and find our results to agree within errorbars. The behavior of the energy is found
    to be well described by $E \propto T^2$, implying a $T$-linear specific 
    heat $C$.}
    \label{fig:energyconvergence}
\end{figure}

Furthermore, we have performed an analysis of the effects of 
finite-bond dimension $D_{\max}$ for our results on the energy $E$ as a function of temperature, as shown in \cref{fig:energyconvergence}. We find 
the energy as a function of temperture exhibits $E \propto T^2$ behavior in
the regime where $0.0125 \leq T/t \leq 0.4$, which implies that the specific heat $C$ is approximately linear in this regime. 

\section{Comparison of METTS cylinder geometries}
\label{sec:geometries}

\begin{figure}[t]
    \centering
    \includegraphics[width=\columnwidth]{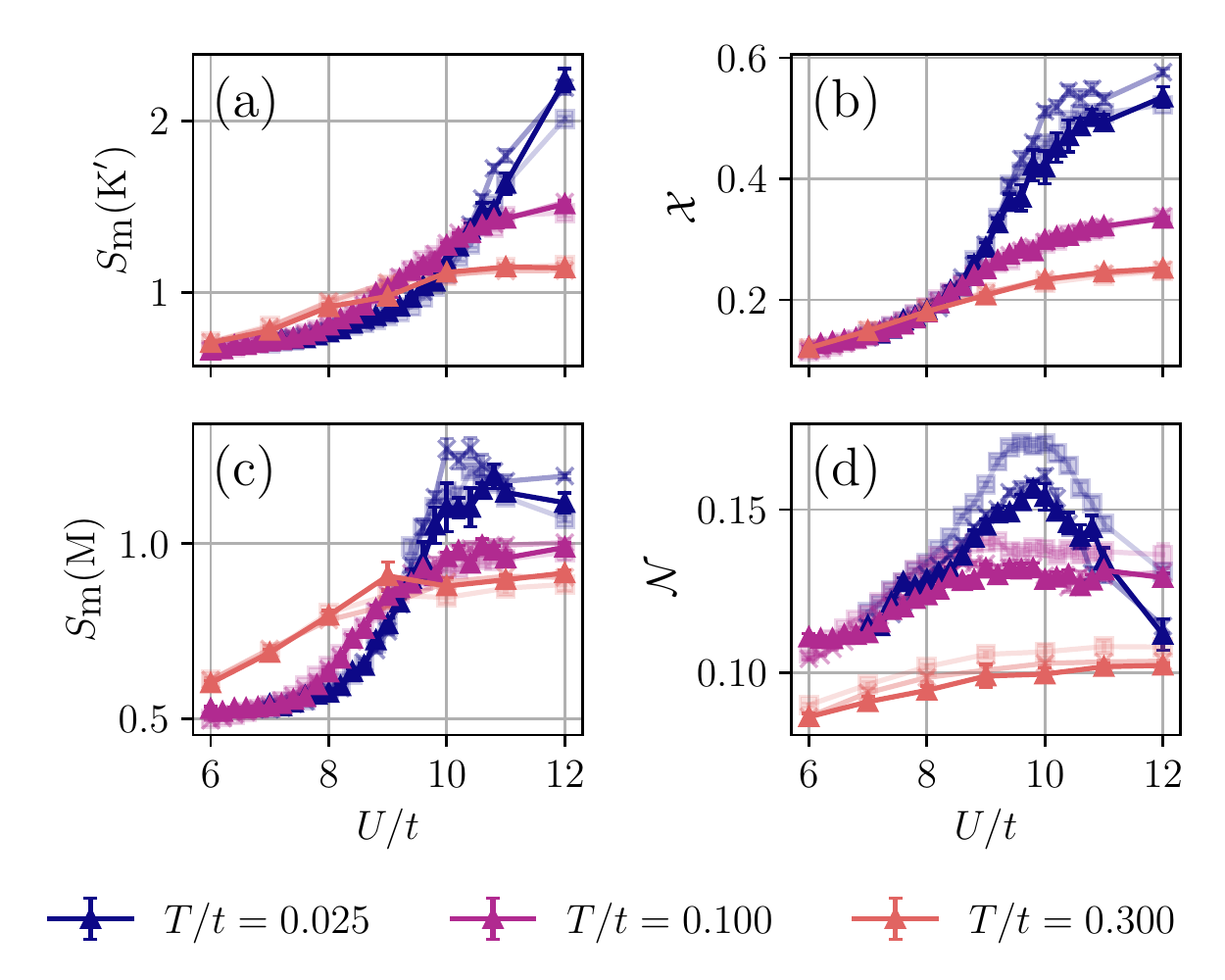}
    \caption{Size dependence of magnetic structure factor
    $S_{\textrm{m}}(\mathbf{k})$ and the chiral susceptibility $\mathcal{X}$ 
    for three different temperatures. We compare  YC4 cylinders of length 
    $L=16$ (squares), $L=24$ (crosses), and $L=32$ (triangles). Increasing opacity denotes longer cylinder length. METTS
    simulations have been performed with a maximal bond dimension
    $D_{\max}=4000$ (a) Magnetic structure
    factor $S_{\textrm{m}}(\textrm{K}^\prime)$ (b) Chiral susceptibility 
    $\mathcal{X}$ (c) Magnetic structure factor $S_{\textrm{m}}(\textrm{M})$ 
    (d) Nematic spin correlation $\mathcal{N}$, from \cref{eq:nematic}.
    We observe only weak dependence of these observables on the length $L$ 
    of the cylinder.
    }
    \label{fig:magnetism_size}
\end{figure}

The results from METTS in the main text have mainly been obtained
on the $16\times 4$ YC4 cylindrical geometry, which is shown in
\cref{fig:snapshots}. In this appendix, we discuss the effects of the cylinder
geometry on our results. First, we investigate the dependence of the magnetic
observables on the cylinder length $L$ of the YC4 cylinder in
\cref{fig:magnetism_size}. We compare simulations at temperatures
$T/t=0.025, 0.100, 0.300$ for $6\leq U/t \leq 12$. We find that our results only
weakly depend on the cylinder length $L$ and are, therefore, expected to be
robust in the limit $L\rightarrow\infty$ for the YC4 cylinder. 

\begin{figure}[t]
    \centering
    \includegraphics[width=\columnwidth]{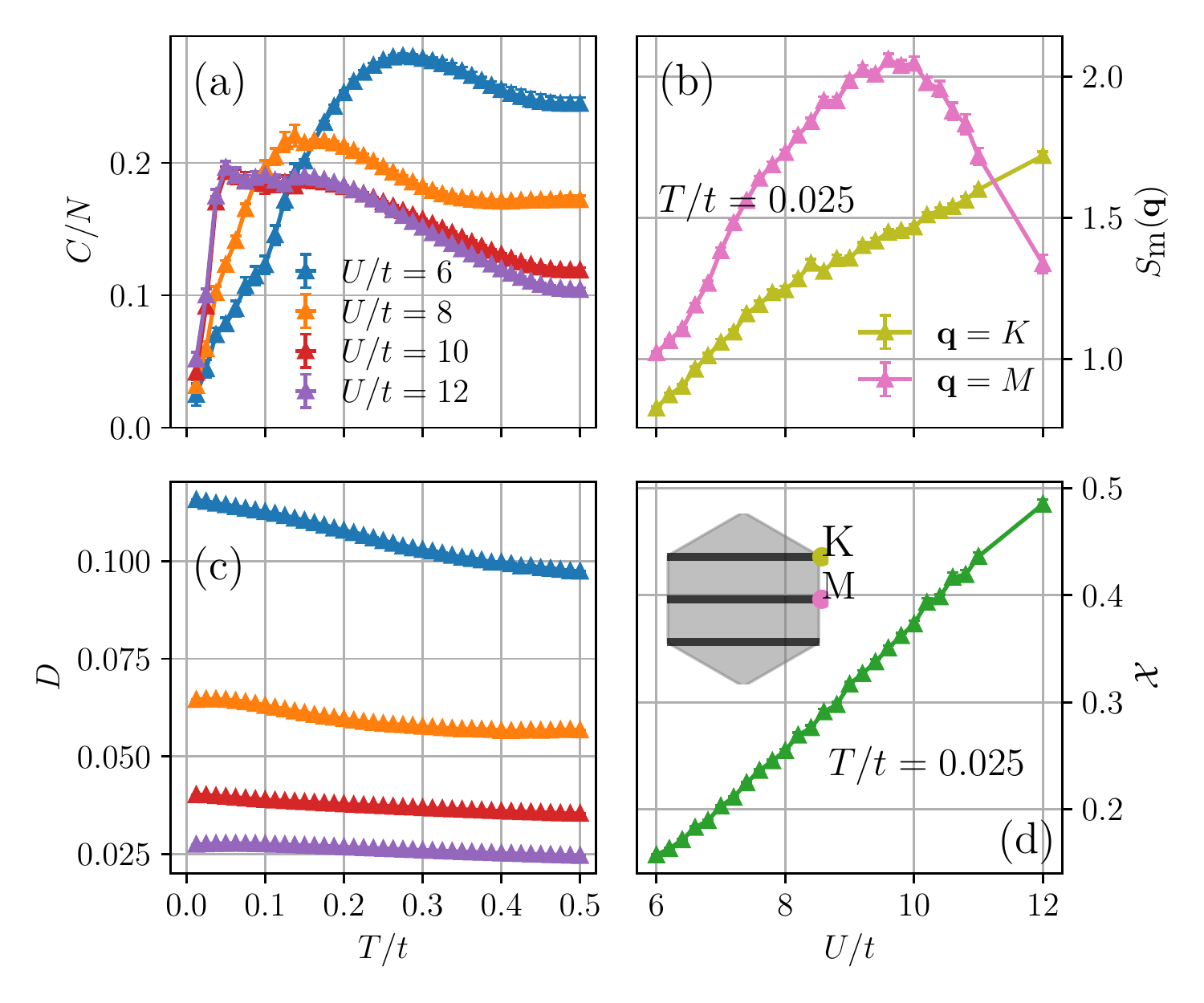}
    \caption{Key results on the YC3 cylinder from METTS. (a) Specific heat $C$
    exhibiting a small peak at $T/t\approx 0.05$ in the intermediate and strong
    coupling regime. (b) Magnetic structure factor $S_{\textrm{m}}(\mathbf{q})$
    at $T/t=0.05$ for $\mathbf{q}=\textrm{M}$ and $\mathbf{q}=\textrm{K}$. We
    observe a peak at $\mathbf{q}=\textrm{M}$ in the intermediate coupling regime.
    (c) The double occupancy $D$ decreases as a function of temperature. 
    (d) The chiral susceptibility smoothly increases as a function of $U/t$.}
    \label{fig:cylindersYC3}
\end{figure}

\begin{figure}[t]
    \centering
    \includegraphics[width=\columnwidth]{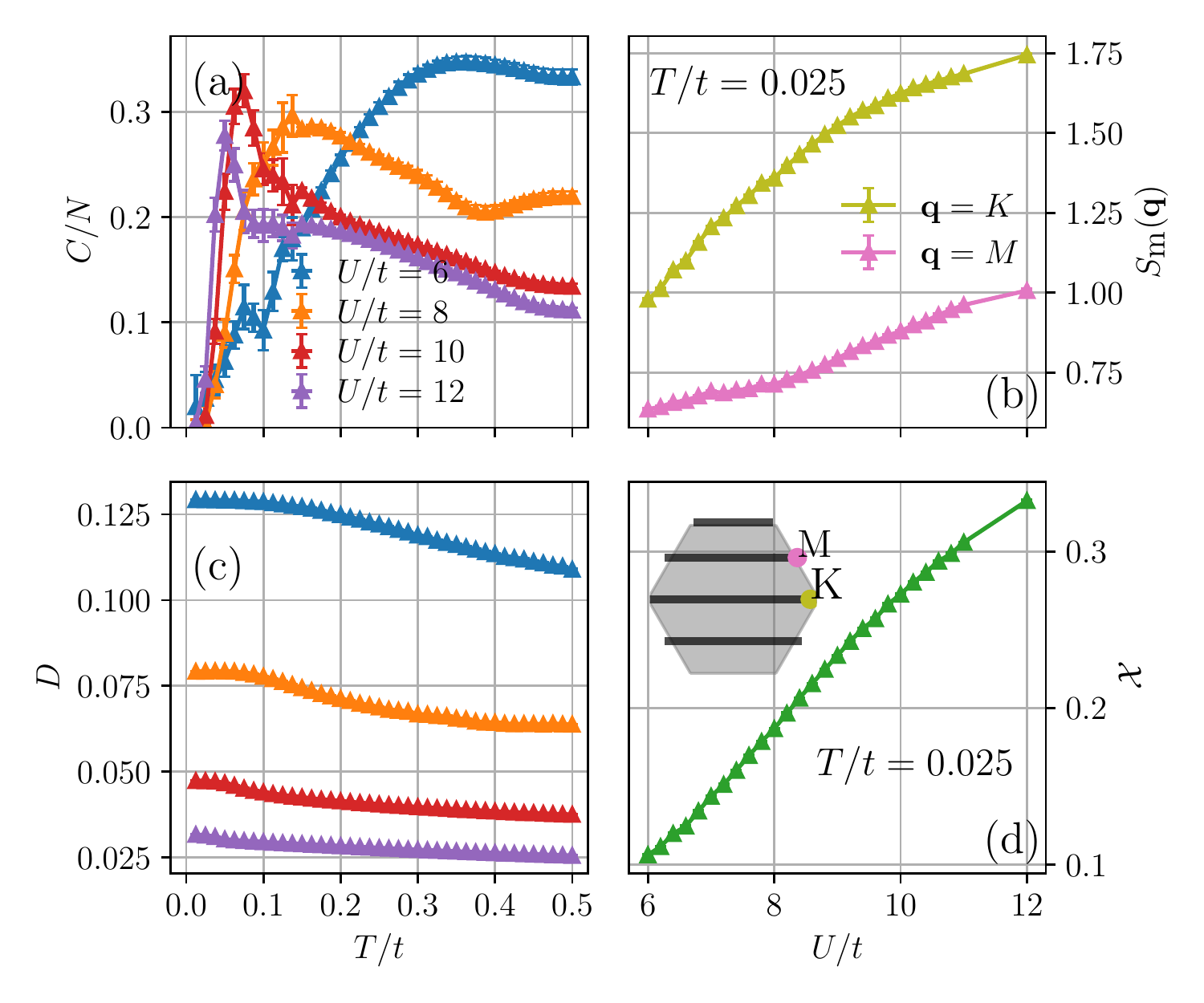}
    \caption{Key results on the XC4 cylinder from METTS. (a) Specific heat $C$ exhibiting a small peak at $T/t\approx 0.05$ in the intermediate and strong coupling regime. (b) Magnetic structure factor $S_{\textrm{m}}(\mathbf{q})$
    at $T/t=0.05$ for $\mathbf{q}=\textrm{M}$ and $\mathbf{q}=\textrm{K}$. We do not observe an intermediate peak $\mathbf{q}=\textrm{M}$. (c) The double
    occupancy $D$ decreases as a function of temperature. (d) The chiral susceptibility smoothly increases as a function of $U/t$.}
    \label{fig:cylindersXC4}
\end{figure}

Moreover, we assess the effect of the cylinder width and boundary conditions on
our results. Here, the differences between the geometries are more pronounced,
and the physics at lowest temperatures in the intermediate coupling regime differs
in some aspects. This has previously already been observed in
Refs.~\cite{Szasz2020,Szasz2021}, where a detailed comparison of DMRG results on
different geometries has been performed. We focus on the YC3 and XC4 geometries.
The YC3 geometry is similar to the YC4 geometry shown in \cref{fig:snapshots}, 
but has a circumference of $L_y=3$. The YC3 geometry allows for stabilizing the
$120^\circ$ N\'{e}el order, and features both the K and M points in reciprocal space. The XC4 geometry, on the other hand, has a circumference of $L_y=4$, but differs from the YC4 geometry by having a distinct periodicity vector given by 
$T=(0, \frac{\sqrt{3}L_y}{2})$. The $120^\circ$ N\'{e}el order is unfrustrated
on this lattice and both K and M points are featured in reciprocal space. 
The resolved momenta of the YC3 and XC4 geometry are shown in \cref{fig:cylindersYC3} and \cref{fig:cylindersXC4}, respectively.

We show the behavior of the specific heat $C$, magnetic structure factor
$S_{\textrm{m}}(\mathbf{q})$, double occupancy $D$,
and chiral susceptibility $\mathcal{X}$ for the YC3 cylinder in \cref{fig:cylindersYC3}(a) and the XC4 cylinder in \cref{fig:cylindersXC4}(a).
First, the behavior of the specific heat is similar on all cylinder geometries 
we investigated. At $U/t=6$ we observe a broad maximum at $T/t \approx 0.25 - 0.4$.
This maximum is shifted towards lower temperatures of around $T/t \approx 0.15$
at $U/t=8$. In the intermediate to strong coupling regime, we observe a small maximum
at temperatures around $T/t=0.05$ followed by an extended plateau. 

At low temperatures, we observe an increase of the specific heat as a function of 
$U/t$, which by \cref{eq:thermalentropy} implies an increase in entropy upon
increasing $U/t$. As discussed in \cref{sec:thermodynamics}, this increase 
in entropy is related to a decrease of the double occupancy $D$ with temperature
via the Maxwell relation \cref{eq:maxwellrelation}. The double occupancy $D$ 
for the YC3 and XC4 geometries is shown in \cref{fig:cylindersYC3}(c) and
\cref{fig:cylindersXC4}(c). The decrease in double occupancy is clearly observed 
in all geometries. Hence, the Pomeranchuk effect is consistently realized on all
geometries we investigated. 

The magnetic structure factor $S_{\textrm{m}}(\mathbf{q})$ at temperature
$T/t=0.025$ as a function of $U/t$ is shown in \cref{fig:cylindersYC3}(b) for 
the YC3 cylinder and \cref{fig:cylindersXC4}(b) for the XC4 cylinder. The YC3
cylinder exhibits a clearly pronounced peak at $\mathbf{q}=\textrm{M}$ in the 
intermediate coupling regime. This is consistent with our results on the YC4
cylinder, where we similarly detected stripy antiferromagnetic order. Also, 
at strong coupling the structure factor is peaked at $\mathbf{q}=\textrm{K}$,
indicating $120^\circ$ N\'{e}el order. However, we do not observe a pronounced
chiral susceptibility $\mathcal{X}$ in the intermediate coupling regime in
\cref{fig:cylindersYC3}. Instead, the chiral susceptibility smoothly increases
as a function of $U/t$. This is consistent with the ground state DMRG study
performed in Ref.~\cite{Szasz2020}, where no chiral spin liquid has been
observed for periodic boundary conditions on the YC3 cylinder. 

Similarly, the XC4 cylinder also does not exhibit a pronounced chiral susceptibility
in the intermediate coupling regime in \cref{fig:cylindersXC4}(d). Also for this
geometry, Ref.~\cite{Szasz2020} reported the chiral spin liquid not being realized
at $T=0$ for periodic boundary conditions. In contrast to the YC3 and YC4
geometries, the XC4 cylinder does not exhibit a pronounced peak of the magnetic
structure factor at the M point in the intermediate regime, as shown in
\cref{fig:cylindersXC4}. Instead, the peak at $\mathbf{q}=\textrm{K}$ is
smoothly increasing as a function uf $U/t$. This demonstrates that the precise 
nature of the state realized in the intermediate regime at low-temperatures
is dependent on the particular cylinder geometry. We think, the kind of
order being exactly realized in the two full-dimensional limit is still to be
determined. However, the pronounced the stripy antiferromagnetic correlations on
the YC3 and YC4 cylinder and the evidence for a CSL on the YC4
cylinder~\cite{Szasz2020,Chen2021} show that these two kinds of orderings are competing
at the lowest temperatures, and might be realized in the full two-dimensional
limit, possibly upon adding further interaction terms.

\section{Magnetic phase transition in (dynamical) mean-field theory}
\label{app:dmft}

\begin{figure}[t!]
    \centering
    \includegraphics[width=\columnwidth]{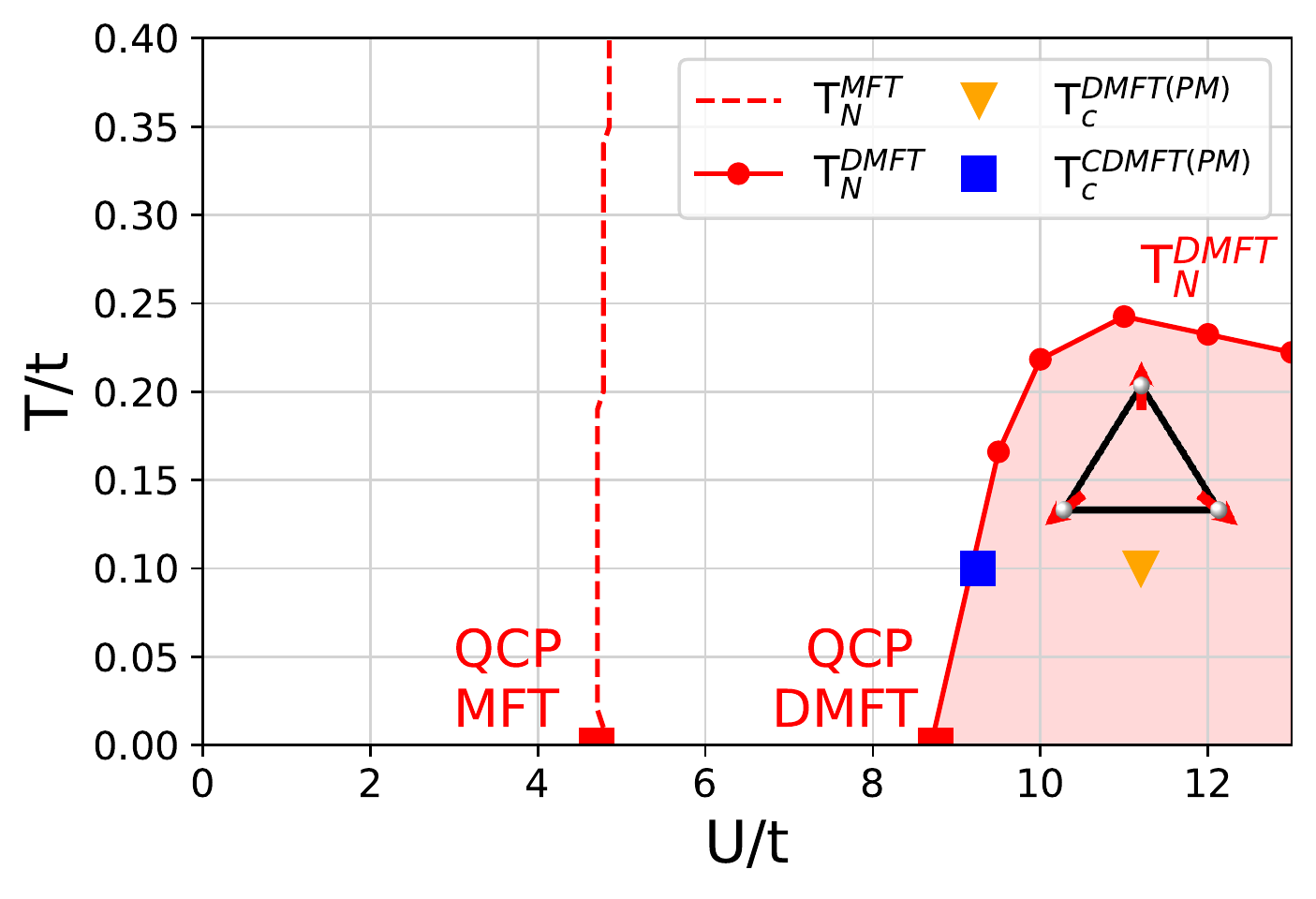}
    \caption{Magnetic phase diagram of the isotropic triangular Hubbard model calculated by MFT (red dashed line) and DMFT (red circles and solid line). The red lines mark the second order transition from a paramagnet (white) to a $120^\circ$ N{\' e}el ordered phase (red shaded). Also shown are the critical temperatures and interactions of the (PM restricted) DMFT (orange triangle) and CDMFT (blue square). Please note that the CDMFT solution has been restricted to its paramagnetic phase. The DMFT zero temperature point has been taken from the ground state calculation of  \cite{Goto2016}.}
    \label{fig:dmft_magnetic}
\end{figure}

\begin{figure}[t!]
    \centering
    \includegraphics[width=\columnwidth]{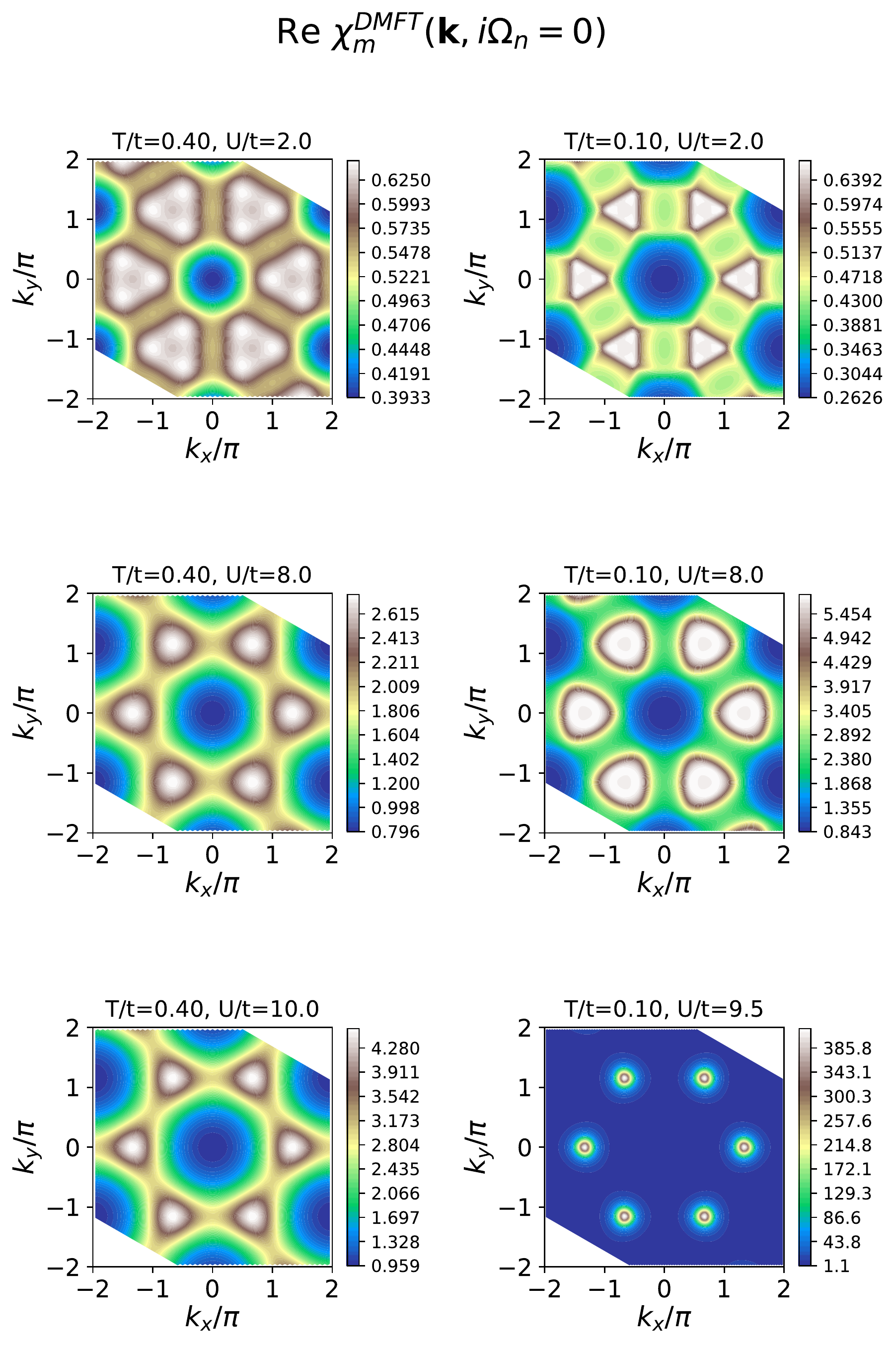}
    \caption{Momentum dependence of the DMFT magnetic susceptibility at zero frequency for several temperatures and interactions. The leading contribution is (centered around) $\mathbf{k}\!=\!K$.}
    \label{fig:chi_K_DMFT}
\end{figure}

\begin{figure}[t!]
    \centering
    \includegraphics[width=0.8\columnwidth]{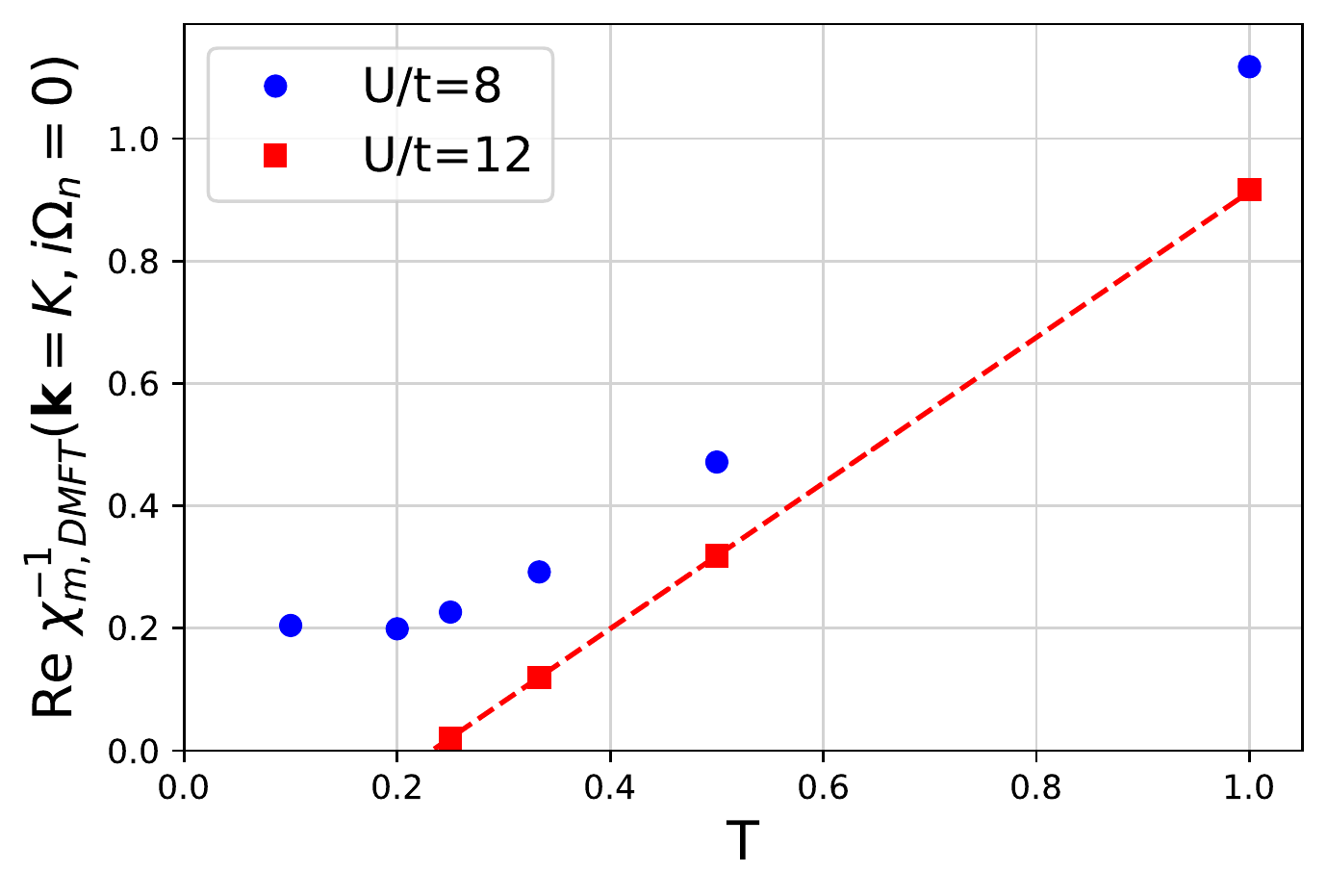}
    \caption{Temperature dependence of the (inverse) magnetic susceptibility at $\mathbf{k}=K$ and zero Matsubara frequency for two different interaction values calculated by DMFT. For $U/t\!=\!12$ the linear fit to determine $T_N^\text{DMFT}$ is shown.}
    \label{fig:chi_inv_DMFT}
\end{figure}

In this Appendix we give an overview of the magnetic properties of the Hubbard model on the isotropic triangular lattice calculated by means of the dynamical mean-field theory (DMFT). By the inclusion of all temporal correlations present in the Hubbard model Eq.~(\ref{eq:hubbardmodel}), DMFT has proven to provide a good starting point for the application of more sophisticated techniques, which aim at including spatial correlations on top (see, e.g., \cite{Schaefer2020,Li2014,Laubach2015,Li2020}).

The main panel of Fig.~\ref{fig:dmft_magnetic} shows the N{\'e}el temperature calculated in DMFT $T_N^{\text{DMFT}}$ (red line and circles) and static mean-field theory (MFT, dashed line), separating the paramagnetic from a magnetically ordered phase \footnote{As a spatial mean-field theory, both MFT and DMFT exhibit a  second order phase transition (for the exact solution the Mermin-Wagner theorem \cite{Mermin1966, Hohenberg1967} prohibits magnetic ordering at finite $T$).}. In contrast to the case of the Hubbard model on a square lattice where due to the nesting properties of the Fermi surface the order appears for every finite $U$ at low enough temperatures, here a quantum critical point separates a Fermi liquid from a magnetically ordered ground state at $U_\text{QCP}^\text{DMFT}/t\!\approx\!9.5$ (cf. also \cite{Li2014,Laubach2015}). From there $T_N^{\text{DMFT}}$ increases steeply with a maximum of $T_{N,\text{max}}^{\text{DMFT}}/t\approx 0.25$ around $U/t\!=\!11$ before slowly decreasing again.

Two points are particularly noteworthy:
\begin{enumerate}
    \item[(i)] as in the case of the Hubbard model on a square lattice the critical end point of the Mott MIT (orange triangle) visible in the paramagnetically restricted DMFT is {\it shadowed}, i.e. preempted, by the magnetic phase transition of DMFT (with the magnetically order phase being the thermodynamically stable phase of DMFT) and
    \item[(ii)] the CDMFT critical end point (blue square) lies close to the phase boundary of the magnetic phase. Please note that we did not calculate the magnetic phase diagram in case of CDMFT, which is restricted to its paramagnetic solution.
\end{enumerate}
Comparing the magnetic phase diagram of DMFT to the self-energies of CDMFT at $T/t\!=\!0.10$ presented in Fig.~\ref{fig:self_energy} of the main text, one can observe that the nearest-neighbor component of the CDMFT self-energy starts to increase at the interaction value $U/t\!=\!9.5$, where the DMFT orders magnetically. In other words the spatial mean-field approximation reflects the increase of non-local fluctuations by entering an ordered phase.

For the determination of the DMFT phase boundary, we calculated the momentum-dependent magnetic susceptibility $\chi_{m}^{\text{DMFT}}(\mathbf{k}, i\Omega_n=0)$ at zero Matsubara frequency by means of the solution of the Bethe-Salpeter equations with the irreducible vertex extracted from the self-consistently determined Anderson impurity model \cite{Georges1996}, using the continuous time quantum Monte Carlo solver in its interaction expansion (CT-INT) and the tprf framework \cite{tprf} of TRIQS \cite{TRIQS}. For the vertex we used up to $N_{i\omega}=50$ positive fermionic Matsubara frequencies and extrapolated the value of the physical susceptibility to $N_{i\omega}\rightarrow\infty$ with $\chi\sim a + b/N_{i\omega}$ (see, e.g., Supplemental Material of \cite{Schaefer2017}).

\begin{figure}[t!]
    \centering
    \includegraphics[width=0.65\columnwidth]{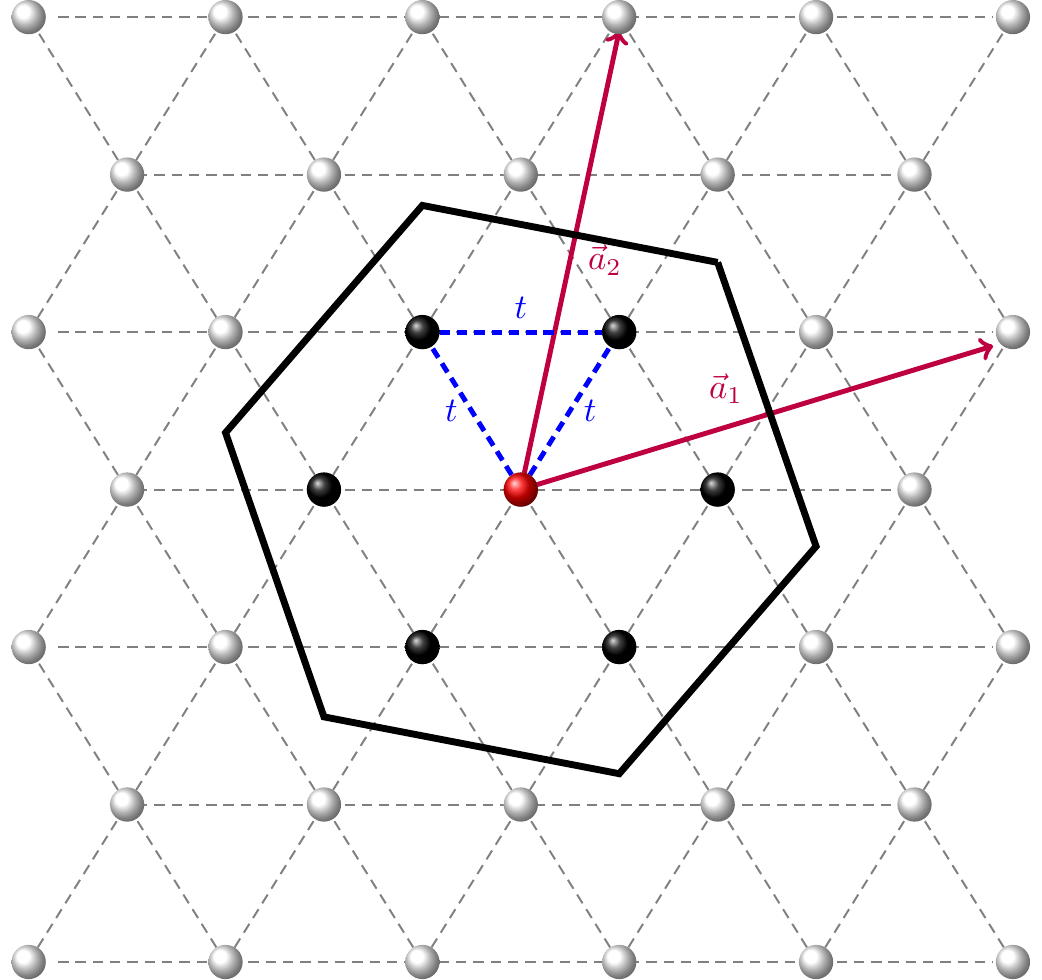}
    \caption{Cluster geometry with $N_c\!=\!7$ used in CDMFT. The cluster consists of a central site (marked in red) and six equivalent outer sites arranged on a ring. The translation vectors are $\vec{a}_1=(5/2, \sqrt{3}/2)$ and $\vec{a}_2=(1/2, 3\sqrt{3}/2)$.}
    \label{fig:cdmft_cluster}
\end{figure}

Due to the second order nature of the phase transition, approaching the phase boundary $\chi_{m}^{\text{DMFT}}(\mathbf{k}, i\Omega_n\!=\!0)$ diverges at the ordering vector $\mathbf{k}\!=\!\mathbf{Q}$. Fig.~\ref{fig:chi_K_DMFT} shows $\chi_{m}^{\text{DMFT}}(\mathbf{k}, i\Omega_n\!=\!0)$ at $T/t\!=\!0.40$ (left column) and $T/t\!=\!0.10$ (right column) for several interaction values. One can see that the leading contribution always stems from momentum vectors centered around $\mathbf{k}\!=\!K$. Approaching the transition, at $T/t\!=\!0.10$, $\chi_{m}^{\text{DMFT}}(\mathbf{k}, i\Omega_n\!=\!0)$ continuously grows before it eventually diverges at $\mathbf{k}\!=\!K$. The temperature dependence of the inverse susceptibility $\chi_{m,\text{DMFT}}^{-1}(\mathbf{k}\!=\!K, i\Omega_n\!=\!0)$ is shown in Fig.~\ref{fig:chi_inv_DMFT} for two different values of the interaction. At $U/t\!=\!8$ (with a Fermi liquid ground state present in DMFT) it exhibits Pauli-like behavior, i.e. approaching a constant at low temperatures. At $U/t\!=\!10$ (with a magnetically ordered ground state) the susceptibility diverges as $\chi\!\sim\!\left|T-T_N^\text{DMFT}\right|^{-\gamma_T}$ at  $T_N^\text{DMFT}/t\!\approx\!0.22$ with $\gamma_T=1$ being the susceptibility's mean-field critical exponent.

\section{Cellular dynamical mean-field theory: cluster geometry, Matsubara data and computational details}
\label{app:cdmft}
\begin{figure}[t!]
    \centering
    \includegraphics[width=0.94\columnwidth]{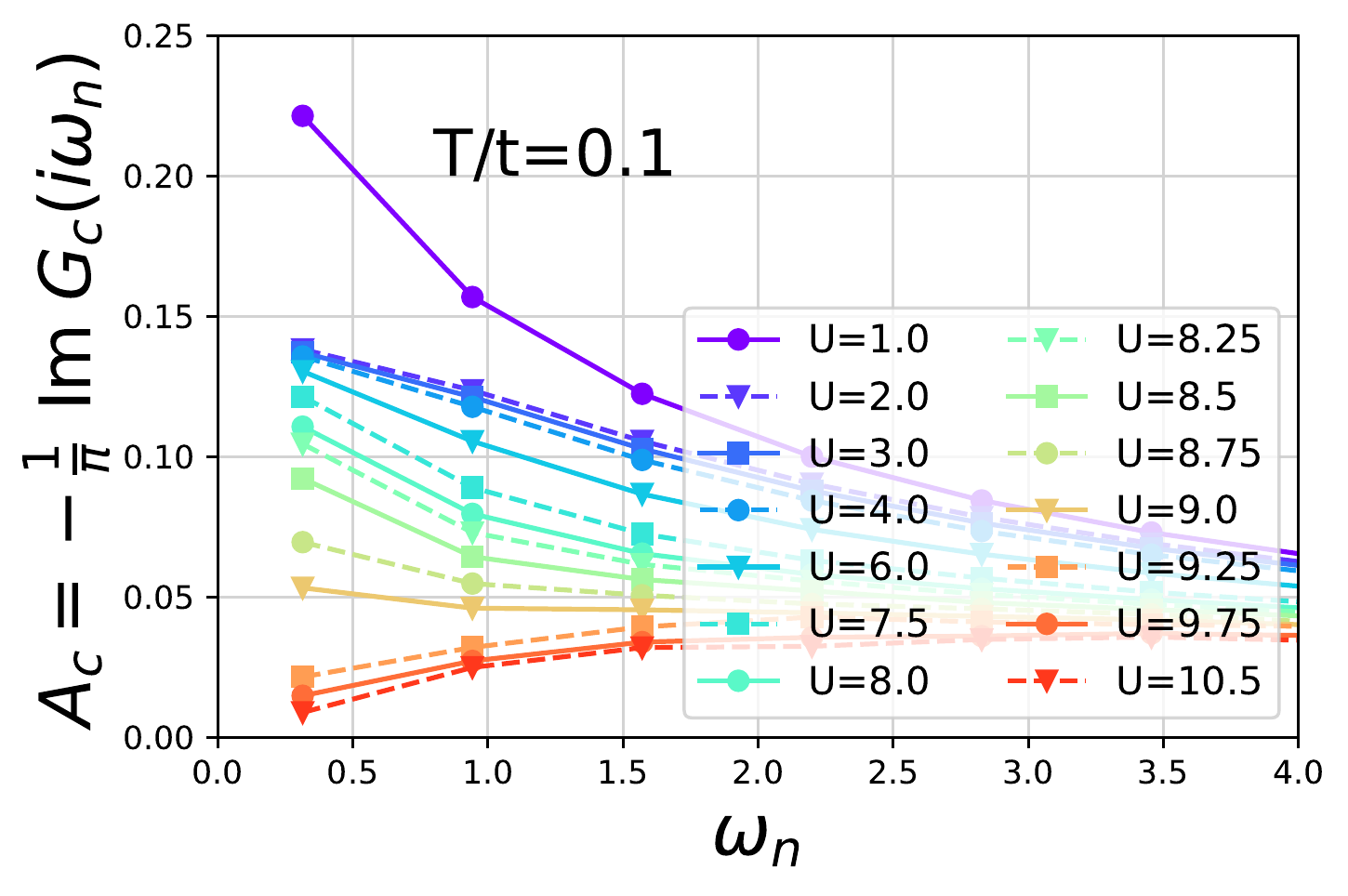}
    \includegraphics[width=0.47\columnwidth]{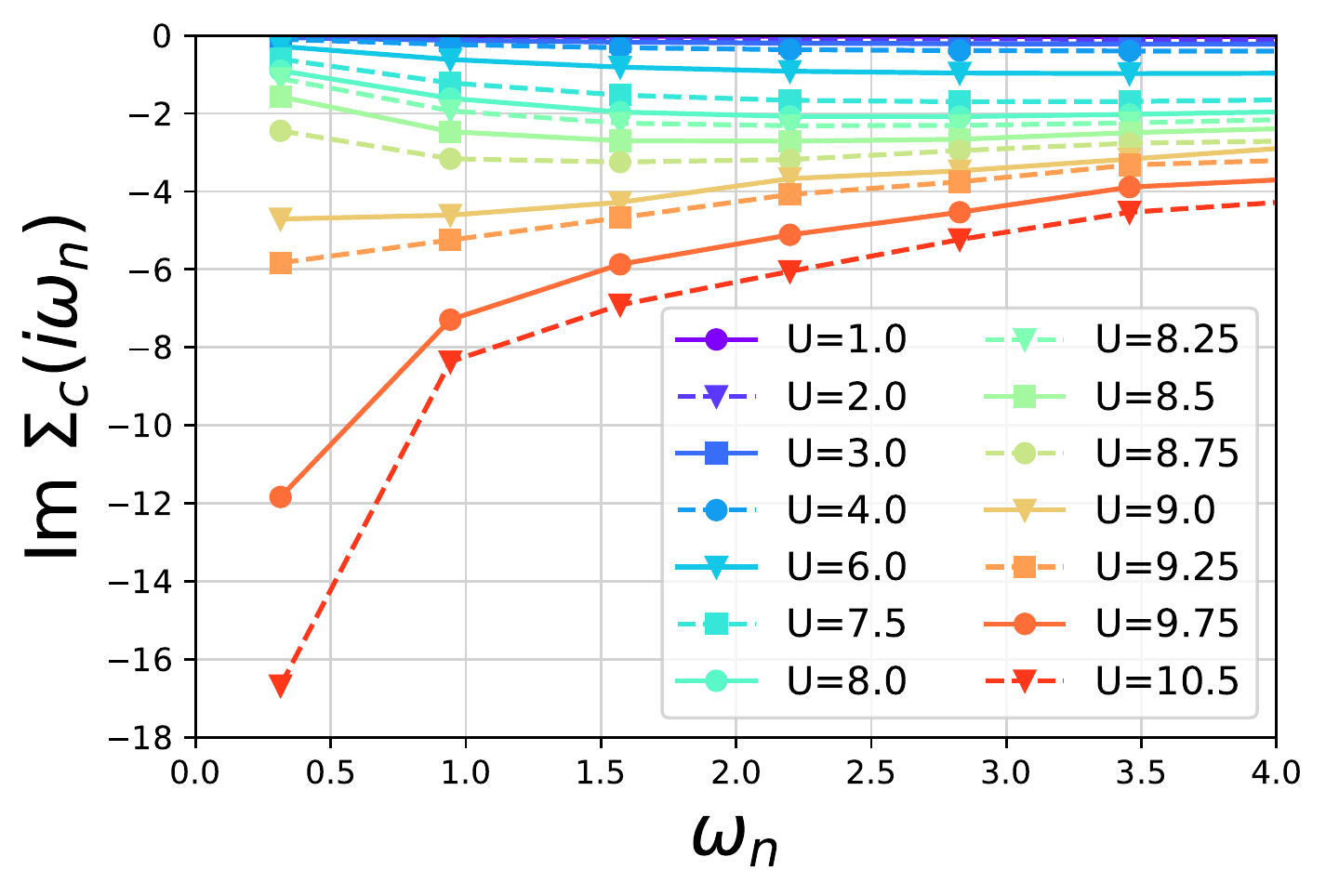}
    \includegraphics[width=0.47\columnwidth]{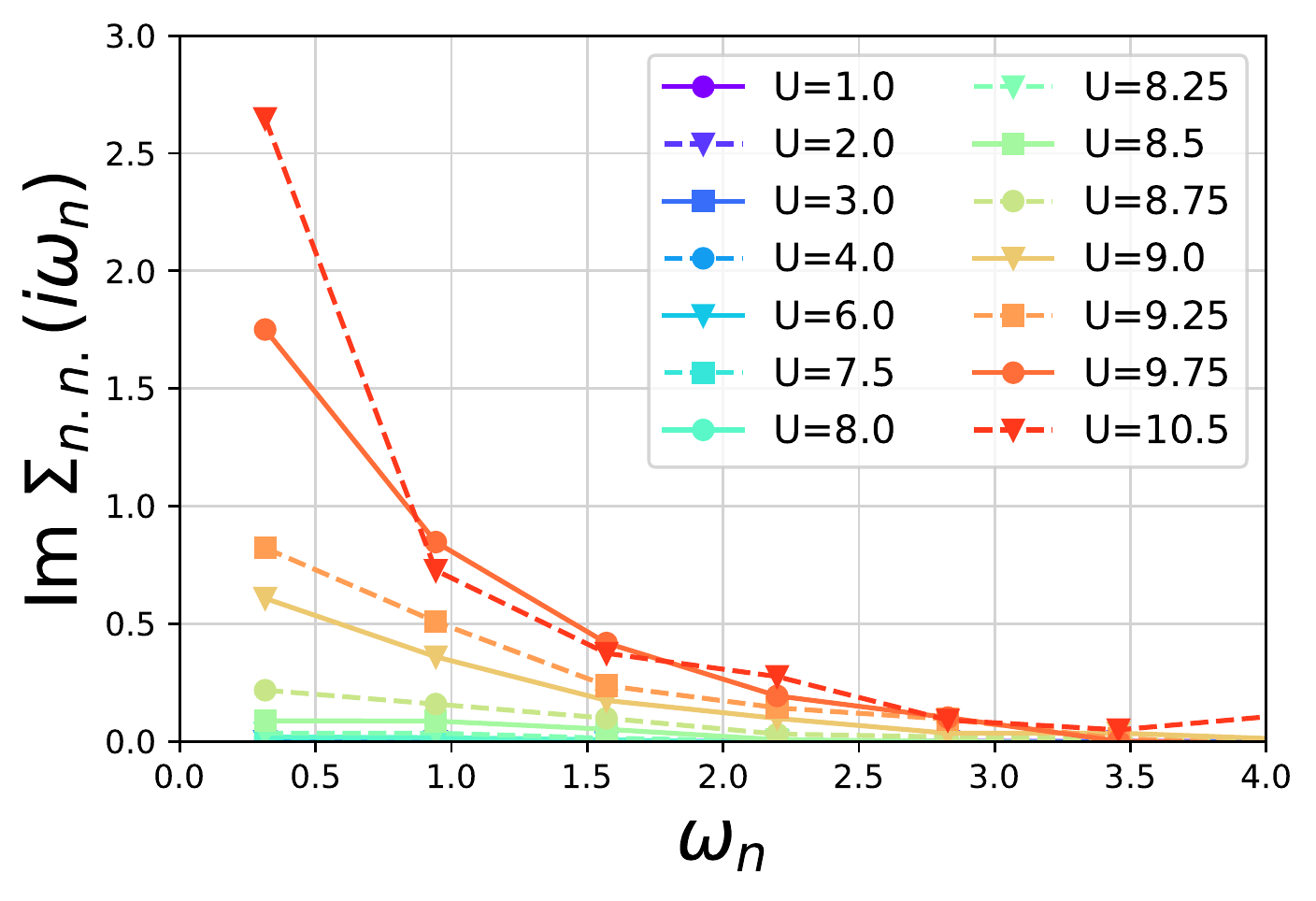}
    \includegraphics[width=0.47\columnwidth]{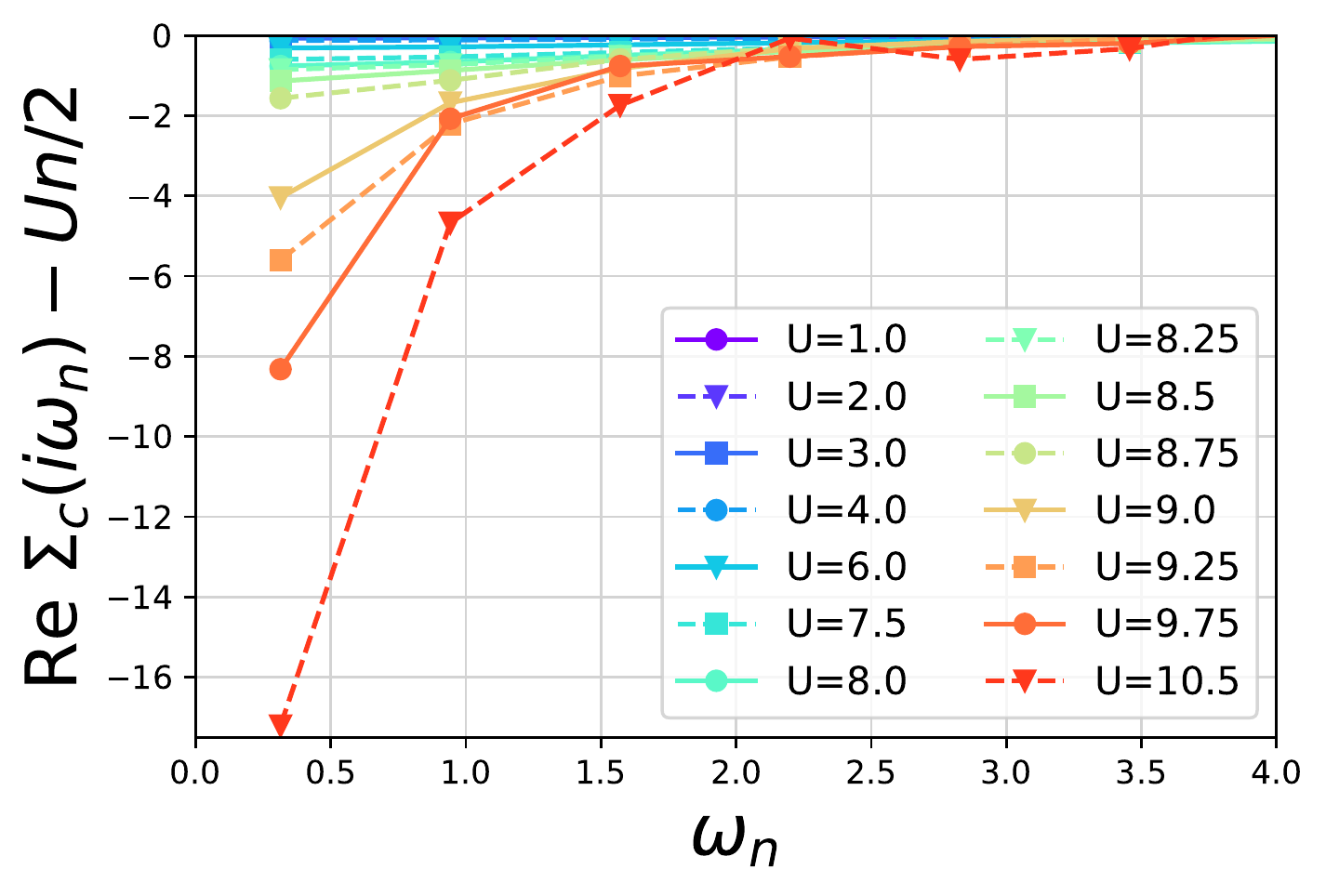}
    \includegraphics[width=0.47\columnwidth]{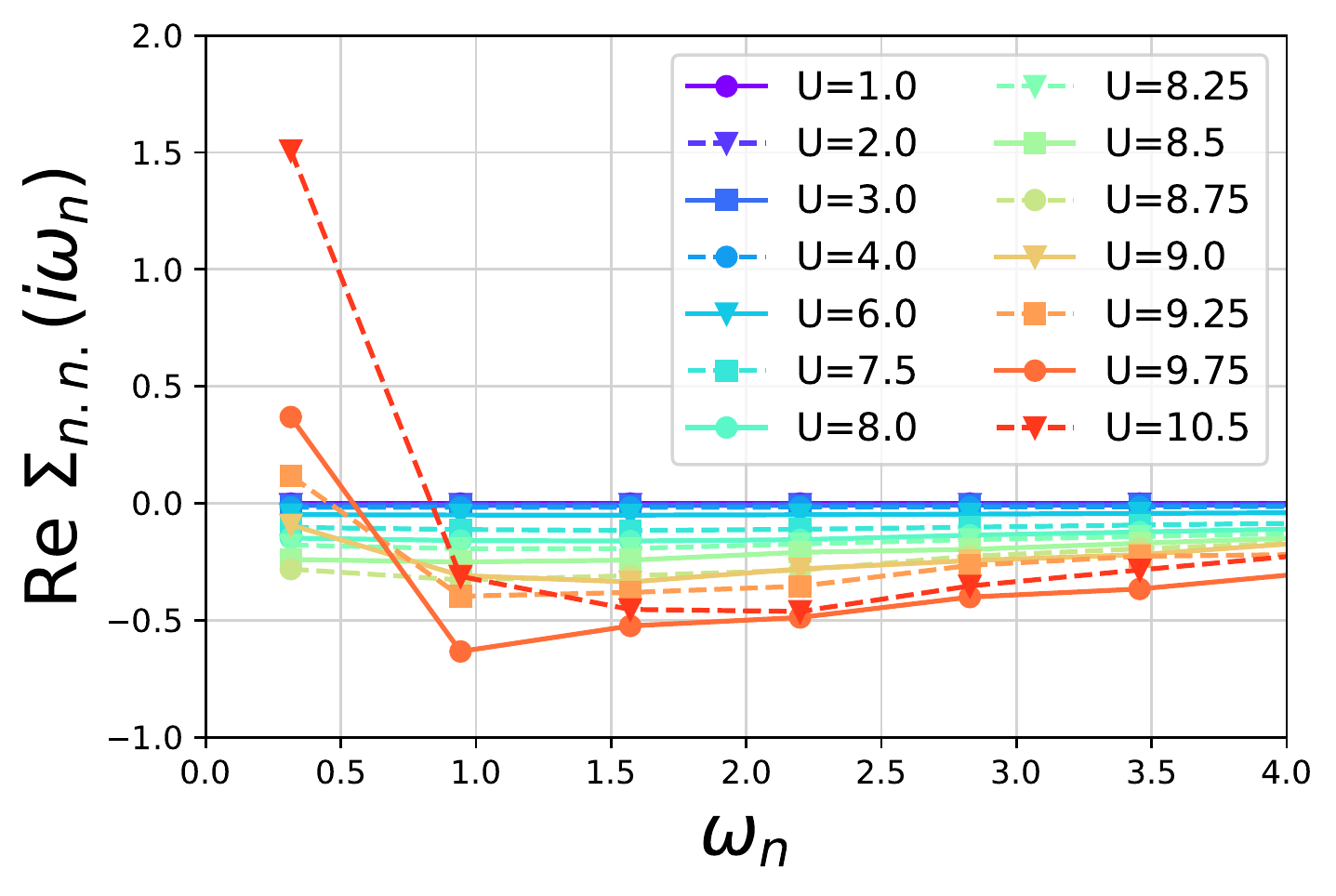}
    \caption{(Upper row) Spectral weight $A_c$ expressed with the imaginary part of the Green function on the central site of CDMFT at $T/t=0.1$ for several interaction values as a function of the Matsubara frequency. (Central row) Imaginary part of the self-energy on the central site (left) and to its nearest neighbor (right). (Lower row) Real part of the self-energy on the central site (left, the Hartree term has been subtracted here) and to its nearest neighbor (right). }
    \label{fig:sigma_CDMFT_iwn}
\end{figure}

\subsubsection{CDMFT on a $N_c\!=\!7$ site cluster, restricted to the PM solution [CDMFT, CDMFT-7 (PM)]}
For the CDMFT calculations performed in this work we used $N_c\!=\!7$ sites which are arranged according to Fig.~\ref{fig:cdmft_cluster} with a central site and six equivalent sites that form an outer ring. We restrict the CDMFT to its paramagnetic solution. Due to the previously found observation \cite{Klett2020} that the self-energy obtained from a cluster center focused extrapolation converges faster with the cluster size than the periodization schemes previously introduced in the literature, for single-particle observables (like the spectral function shown in panel (a) of Fig.~\ref{fig:mit_transition}) and potential energies (double occupancies) we show values for the central site. Similarly for the self-energies shown in Fig.~\ref{fig:self_energy} we took the central site as representative for its local component and as its nearest-neighbor component the values from central site to one of the (equivalent) outer ring sites. This results in a remarkable good agreement with the results from numerically exact DiagMC in the regimes where DiagMC can be controllably resummed.

For completeness and reference in Fig.~\ref{fig:sigma_CDMFT_iwn} we show the Matsubara frequency dependence of the single-particle properties spectral function $A_c$ as expressed by the Green function (upper panel), the imaginary part (central panel) and real part (lower panel) of the self-energy. These quantities are shown for $T/t=0.1$ and for several values of the interaction $U/t$. For the self-energy we show both the central site and nearest neighbor values. For the real part at the central site we have subtracted the respective Hartree term. The MIT is clearly visible between $U/t = 9$ and $9.25$ as (i) a suppression of the spectral weight at low frequencies and (ii) a change of slope \cite{Schaefer2015, Simkovic2020}) and eventually divergence of the imaginary part of the self-energy on the central site with increasing $U/t$. The data of  Fig.~\ref{fig:mit_transition}~(a) in the main have been obtained from a linear extrapolation of the data in the upper panel of Fig.~\ref{fig:sigma_CDMFT_iwn} to zero frequency.

For the calculations of the self-energies, we converged the CDMFT self-consistency cycle using a continuous time quantum Monte Carlo solver in its interaction expansion (CT-INT) in the TRIQS framework \cite{TRIQS}.

\subsubsection{CDMFT on a $N_c\!=\!4$ site cluster, allowing for spin-symmetry breaking (CDMFT-4)}
\begin{figure}[t!]
    \centering
    \includegraphics[width=0.45\columnwidth]{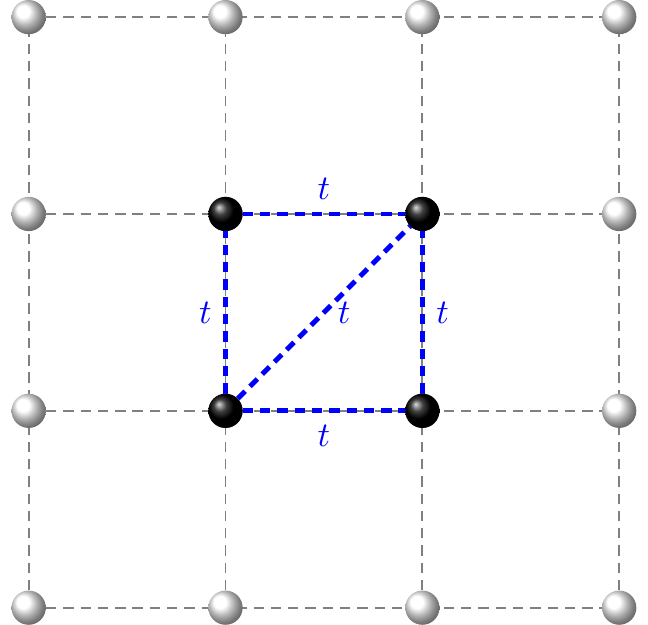}
    \caption{Cluster geometry with $N_c\!=\!4$ used in CDMFT-4.}
    \label{fig:cdmft4_cluster}
\end{figure}

For the CDMFT calculations that allow for spin-symmetry breaking (Fig.~\ref{fig:comparison}, CDMFT-4) we used a $N_c\!=\!4$ cluster with the geometry given in Fig.~\ref{fig:cdmft4_cluster}. As already stated in the main text, the convergence of a larger cluster (at least close to phase transitions) is prohibited by the fermionic sign problem.

For being able to enter the symmetry-broken phase we use an approach similar to the one described in \cite{Goto2016} for single-site DMFT. However, instead of rotating the Green function we rotate an external magnetic field $H\sigma_z$ by applying:
\begin{equation}
    \mathrm{e}^{\mathrm{i} \phi \sigma_y/2} \left[H \sigma_z \right]  \mathrm{e}^{-\mathrm{i} \phi \sigma_y/2}
\end{equation}
where $\sigma_i, i \in \left\{x,y,z\right\}$ denote the Pauli matrices. The $120^\circ$ N\'eel ordering corresponds to a rotation with $\phi=\vec{Q}\vec{R}_i, \vec{Q}=(2\pi/3,2\pi/3)$. The breaking of the $SU(2)$-symmetry of the Green function results in an effective $8$-orbital calculation at the impurity level. After some iterations of the algorithm we switch off the field and let the system converge self-consistently.


\bibliography{hubbard_triangular.bib}

\end{document}